\documentclass[twocolumn]{aastex63}

\usepackage{booktabs}
\usepackage{graphicx}
\usepackage{subfigure}
\usepackage{soul}
\usepackage{tablefootnote}
\usepackage{amsmath}
\usepackage{gensymb}
\usepackage{float}
\usepackage{textcomp}
\usepackage{gensymb}
\usepackage{silence}
\newcommand{\R}[1]{{{#1}}}
\newcommand{\RR}[1]{{{{#1}}}}
\newcommand{\RRR}[1]{{{#1}}}

\begin{document}

\title{Orbital Phase-resolved Analysis of X-ray and Gamma-ray Observations of the High-Mass Gamma-ray Binary 4FGL J1405.1-6119}

\author[0000-0003-3540-2870]{Alexander Lange}
\affiliation{Department of Physics, The George Washington University, 725 21st Street NW, Washington, DC 20052, USA}
\affiliation{X-ray Astrophysics Laboratory, Code 662 NASA Goddard Space Flight Center, Greenbelt Rd., MD 20771, USA}
\affiliation{Center for Space Sciences and Technology, University of Maryland, Baltimore County, 1000 Hilltop Circle, Baltimore, MD 21250, USA}

\author[0000-0002-3396-651X]{Robin H.D. Corbet}
\affiliation{X-ray Astrophysics Laboratory, Code 662 NASA Goddard Space Flight Center, Greenbelt Rd., MD 20771, USA}
\affiliation{Center for Space Sciences and Technology, University of Maryland, Baltimore County, 1000 Hilltop Circle, Baltimore, MD 21250, USA}
\affiliation{CRESST II, USA}

\author[0000-0001-7532-8359]{Joel B. Coley}
\affiliation{CRESST II, USA}
\affiliation{Department of Physics and Astronomy, Howard University, Washington, DC 20059, USA}
\affiliation{Astroparticle Physics Laboratory, Code 661 NASA Goddard Space Flight Center, Greenbelt Rd., MD 20771, USA}

\author[0000-0002-5130-2514]{Guillaume Dubus}
\affiliation{Univ. Grenoble Alpes, CNRS, IPAG, 38000 Grenoble, France}

\author[0000-0002-8548-482X]{Jeremy Hare}
\affiliation{X-ray Astrophysics Laboratory, Code 662 NASA Goddard Space Flight Center, Greenbelt Rd., MD 20771, USA}
\affiliation{Center for Space Sciences and Technology, University of Maryland, Baltimore County, 1000 Hilltop Circle, Baltimore, MD 21250, USA}
\affiliation{The Catholic University of America, 620 Michigan Ave., N.E. Washington, DC 20064, USA}

\author[0000-0002-2413-9301]{Nazma Islam}
\affiliation{X-ray Astrophysics Laboratory, Code 662 NASA Goddard Space Flight Center, Greenbelt Rd., MD 20771, USA}
\affiliation{Center for Space Sciences and Technology, University of Maryland, Baltimore County, 1000 Hilltop Circle, Baltimore, MD 21250, USA}

\author{Jonathan Barnes}
\affiliation{Department of Physics and Astronomy, Howard University, Washington, DC 20059, USA}

\begin{abstract}
We present the results of multi-wavelength observations of the High-Mass Gamma-Ray Binary 4FGL J1405.1-6119. \RR{A pair of joint} XMM-Newton and NuSTAR observations taken in 2019 (sampling the gamma-ray maximum and \RR{X-ray maximum}) \RR{characterize} the emission of soft and hard X-rays.  We find variability of the hydrogen column density \RR{along our line of sight}, $N_{\rm H}$, and \R{photon} index, $\Gamma$, and \RR{find no evidence} of pulsations \RR{in X-rays}.  We also \RR{refine} \R{a new best-fit orbital period \RR{to} $P=13.7157\pm0.0014$\,days,} the first orbital phase-resolved analysis based on \R{nearly 16} years of Fermi--LAT observations of 4FGL J1405.1-6119 and the evolution of the spectral shape as a function of orbital phase. Finally, the X-ray and $\gamma$-ray \R{spectra for the phases sampled in the new X-ray observations} can be interpreted in the framework of the intrabinary shock model, previously \R{applied to} High-Mass Gamma-Ray binaries such as LS 5039.

\end{abstract}


\section{Introduction} \label{sec:intro}
High-Mass Gamma-Ray Binaries (HMGBs) are a rare class of $\gamma$\RR{-}ray sources in the Galaxy, consisting of a compact object, most likely a neutron star, but potentially a black hole, 
and a massive companion star (typically, O or B/Be type), that produce a characteristic spectral peak above \R{$\sim$1\,MeV}. The small population of known HMGBs ($\approx$10) have  
debated emission mechanisms of the X-rays and $\gamma$ rays as well as the nature of the compact objects \R{\citep{dubus_sizing_2017,chernyakova_gamma-ray_2020}}. \R{As such, HMGBs provide excellent opportunities to test models of high-energy emission and particle acceleration.} Currently, there are two prevalent models for the high-energy emission: an intrabinary shock (IBS) model with a neutron star compact object and a jet microquasar model with \RR{a neutron star \R{or} black hole} (for reviews, see \citealt{mirabel_revealing_2012};\citealt{dubus_gamma-ray_2013}).

The IBS model proposes that the $\gamma$-ray emission is \RR{produced} by the interactions between the relativistic outflow of a rotation-powered pulsar (i.e., pulsar wind) and the wind of its companion star via \RR{the} inverse Compton scattering of seed photons by the accelerated charged particles. 
In contrast, the microquasar model describes the production of $\gamma$ rays from particles accelerated in a relativistic jet. The formation of this jet is related to accretion onto the compact object (neutron star or black hole).

Among HMGBs, the search for (and subsequent detection of) radio pulsations has identified the compact objects of PSR B1259-63 \R{\citep{johnston_psr_1992}, PSR J2032+4127 \citep{camilo_radio_2009} and LS I +61\degree 303 \citep{weng_radio_2022} as pulsars, with respective spin periods of 47\,ms, 143\,ms, and 269\,ms, supporting the IBS model}. However, the companion's stellar wind can suppress radio emission from a putative pulsar due to free-free absorption such as in \citet{weng_radio_2022}, making it hard to confirm the nature of the other HMGBs' compact objects with only radio-based observations.

The produced $\gamma$ rays may occur in tandem with multi-wavelength flaring and can correspond to certain orbital phases \R{(e.g., X-ray/GeV/TeV, PSR B1259-63; or Radio/GeV, Cygnus X--3; \citealt{chernyakova_multi-wavelength_2015,chernyakova_new_2020}, \citealt{piano_agile_2012})}, but these correlations are not consistent throughout all HMGBs. Beyond the \R{high-energy (HE) - very high-energy (VHE)} $\gamma$-ray regime, HMGBs exhibit modulated emission of X-rays\RR{. This}  places limits on the broadband spectral energy distribution (SED), directly probes the relativistic lepton population, and allows for physical modeling of the IBS and \RR{high-energy} emission \citep{dubus_gamma-ray_2013}. Observed X-ray modulation can be attributed to the changes in geometry of the binary and of the IBS \R{(e.g., PSR B1259-63; \citealt{chernyakova_new_2020})}. The modulation of emission and absorption lines of the companion's spectra can further constrain the orbital solution \R{\citep[see e.g. 1FGL J1018.6-5856;][]{The_Fermi_LAT_Collaboration_2012, 2015ApJ...813L..26S, vansoelen_improved_2022}}. X-ray spectral fitting can reveal spectral breaks and provide estimates of the magnetic field strength at the source of the high-energy emission. Lastly, multi-wavelength observations provide an opportunity to test \RR{these emission models, and determine} the compact object's nature (e.g. HESS J0632+057; \citealt{kim_investigation_2022}). 

The HMGB 4FGL J1405.1-6119 (hereafter referred to as J1405) \R{was discovered by \citealt{corbet_discovery_2019} in a blind periodicity search of {\it Fermi}--LAT fluxes with an orbital period of $13.7157\pm0.0014$\,days. Fermi--LAT probability-weighted aperture photometry data showed an orbital modulation of the $\gamma$ rays with two maxima per orbital cycle.  From the maximum $\gamma$-ray flux and first peak (defined at $\phi=0.0$), there is a sharp drop in flux of the binary followed by a small increase to the second peak ($\phi\approx0.4$) with a slow decay to a minimum. Both peaks are roughly one day in width (corresponding to $\sim$1/14 in phase) where the peak at $\phi=0.0$ has an observed flux of roughly twice that of the peak at $\phi\approx0.4$. There was no spectral analysis beyond fitting the normalization with spectral parameters fixed to the 4FGL catalog \citep{abdollahi_fermi_2020} values, from which J1405 was found to have a spectral peak around 500\,MeV.} 
\R{Follow-up radio measurements from the Australia Telescope Compact Array at 5.5 and 9\,GHz and soft (0.2--10\,keV) X-ray observations from the X-ray telescope aboard the Neil Gehrels Swift Observatory (Swift--XRT) show a coincident source with modulation on the same orbital period. The radio and X-ray \RR{observations} feature a single peak slightly offset from the 2nd $\gamma$-ray peak at $\phi\approx0.3$ and $\phi\approx0.59$, respectively. It should be noted that the modulation of the 5.5\,GHz flux is weaker than at 9\,GHz. There is no X-ray peak at either of the gamma-ray peaks. Lastly, near infrared spectral measurements by FLAMINGOS-2 
classify J1405's companion star as an O6.5 III star \citep{corbet_discovery_2019}.}

The Galactic column density of neutral hydrogen and the photon index were reported as $\mathrm{n}_\mathrm{H} = 6.9\times10^{22}$ cm$^{-2}$ and $\Gamma_{X}= 1.5\pm0.4$ \citep{corbet_discovery_2019}, respectively.

We report on the results of the follow-up \emph{XMM-Newton} and \emph{NuSTAR} observations, and provide an update to the Fermi--LAT analysis to include a phase-resolved analysis of 16\,years of data. We fit the X-ray and $\gamma$-ray data with the IBS model and report the results. The remainder of this paper is structured as follows: the details of each observation and their reduction are reported in Section~\ref{sec:Observations} and the spectral and timing analyses of the X-ray data are reported in Section~\ref{sec:X-ray}. Section~\ref{sec:gamma-ray} provides the results of the timing and orbital phase-resolved spectral analysis of $\sim$16\,years of Fermi--LAT data. Section~\ref{sec:broadband} provides the details of theoretical models used and the modeling of the phase-resolved broadband SEDs. The implications of these results and how they compare to previous studies are discussed in Section \ref{sec:Disc} followed by the conclusion in Section \ref{sec:Conc}.

\section{X-ray and Gamma-ray Observations}
\label{sec:Observations}
Joint XMM-Newton/NuSTAR observations of J1405 \R{were conducted around the Chandra source found at the position (RA, Dec)
= (14$^{\rm h}$05$^{\rm m}$14\fs43, -61\degree 18$'$ 28\farcs36)} in August of 2019 to capture the \R{X-ray behavior at the} $\gamma$-ray maximum and minimum \R{as reported in \citet{corbet_discovery_2019}. We hereafter refer to the observations performed at the $\gamma$-ray maximum and minimum (where the X-ray flux is at a maximum) as ``Gamma-ray Maximum'' \R{($0.964 < \phi_{\mathrm{Gamma{\text -}ray\,Max.}} < 0.0357$) and ``X-ray Maximum'' ($0.393 < \phi_{\mathrm{X{\text -}ray\,Max.}} <
\RR{0.464}$).}} \R{All analyses use a refined binary orbital period of P=13.7157\,days (see Section~\ref{sec:gamma-ray}). A reconstructed X-ray phase-folded lightcurve is shown} in Figure~\ref{fig:1 XrayLC} to compare to Figures~3 and 4 from \citet{corbet_discovery_2019}. The details of the 2019 NuSTAR and XMM-Newton observations are provided in Table~\ref{tab:observations}. \RR{Fermi--LAT photon events spanning 16 years} are used in the Fermi--LAT timing and spectral analyses.

\begin{table*}[htbp]
  \centering
  \caption{4FGL J1405.1-6119 NuSTAR and XMM-Newton Observation IDs, Exposures and Count Rates.}
  \label{tab:observations}
  \begin{tabular}{cccccccccc}
    \toprule
    \textbf{Telescope} & \textbf{Phase Range ($\phi$)} & \textbf{Start Date (UTC)} & \textbf{Observation ID} & \textbf{Exposure (ks)} & \textbf{Count Rate ($\times10^{-2}$ counts s$^{-1}$)}\\
    \midrule
    NuSTAR &  0.91 - 0.05 &2019-08-24 08:24:56& 30502015004 &  86.8  & $1.04 \pm 0.06$\\
    XMM-Newton & 0.93 - 0.96 &2019-08-24 15:01:01& 0852020201 & 49.8 & $4.00 \pm 0.17$\\
    \hline \hline
    NuSTAR &  0.35 - 0.46 & 2019-08-16 16:52:41& 30502015002  & 61.7 & $2.12 \pm 0.09$\\
    XMM-Newton &  0.42 - 0.45 &2019-08-17 16:34:15 & 0852020101 & 33.9 & $3.86 \pm 0.32$ \\
    \bottomrule
  \end{tabular}
\end{table*}

\subsection{XMM-Newton}
The X-ray Multi-Mirror Mission observatory \citep[XMM-Newton,][]{jansen_xmm-newton_2001} consists of three sets of co-aligned X-ray telescopes: two EPIC MOS CCDs and an EPIC PN CCD. EPIC MOS provides coverage between 0.2--10.0\,keV\RR{,} while EPIC PN extends the range to 0.15--15\,keV \citep{turner_european_2001,struder_european_2001}.

Two XMM-Newton observations (50\,ks and 34\,ks, respectively) of J1405 were taken in August 2019. EPIC MOS1/MOS2 cameras observed J1405 in Full Window mode while EPIC PN camera observed in Small Window mode. EPIC MOS has a timing resolution of 0.3 seconds and is not used for timing analysis. Small Window mode allows for a PN timing resolution of 5.7 ms \citep{struder_european_2001} and is used in Section~\ref{sec:X-ray timing}. 

The XMM-Newton observations were reduced using the standard SAS tools version 20.0.0. Both MOS and PN observations are filtered to remove times of high-particle background flaring.

\RR{This filtering reduces the exposures for the Gamma-ray Maximum and X-ray Maximum to 49.8 and 33.9\,ks}. To correct for the difference between the EPIC and NuSTAR effective areas and align the PN spectra to the NuSTAR spectra, we used the \texttt{applyabsfluxcorr=yes}
keyword in \texttt{arfgen}  
as per the latest recommendation by the XMM-SAS team\footnote{XMM-SAS Calibration note: \url{https://xmmweb.esac.esa.int/docs/documents/CAL-TN-0230-1-3.pdf}}.

Circular source and background regions were selected from the same chip around J1405 and a source-free background region. The source and background regions are of radius $r=25''$ for PN and $r=35''$ for MOS1/MOS2. Events filtered from source regions were barycentered, using \texttt{barycen}, and source and background spectra are extracted from their respective regions. Spectral data were grouped into a minimum of 25 counts per bin.
\linebreak
\subsection{NuSTAR}
The Nuclear Spectroscopic Telescope Array \citep[NuSTAR,][]{harrison_nuclear_2013} observatory uses its two modules FPMA and FPMB for 
broad spectral coverage from 3--79\,keV and a timing resolution of 65\,$\mu$s \citep{bachetti_timing_2021}.

NuSTAR performed two observations of J1405 (87\,ks and 62\,ks) contemporaneously with XMM-Newton (see Table~\ref{tab:observations}). \RR{The} NuSTAR observations were reduced using HEASoft version 6.31.1 \citep{nasa_high_energy_astrophysics_science_archive_research_center_heasarc_heasoft_2014}. We used the standard NuSTARDAS pipeline, \texttt{nupipeline} and \texttt{nuproducts} with keywords \texttt{saacalc=2} and \texttt{tentacle=Yes} to account for the South Atlantic Anomaly and any space weather, and generate science products, reducing scientific exposure to 74\,ks and 51\,ks, respectively. 

Source and background regions were chosen from a $r=25''$ circular region and a $r=50''$ circular region around J1405 \R{and around a \RR{region away from the J1405} respectively,} \RRR{on the same detector for both modules.} \R{Variations between the source and background regions are corrected for using \texttt{nubackscale}.} We correct the photon arrival times for barycentric motion and apply the clock correction file ``nuCclock20100101v125.fits'' from the standard CALDB dataset to search for pulsations at high-frequencies detailed in Section~\ref{sec:X-ray timing}. The spectral data are binned to contain a minimum of 25 counts per energy bin. 

\subsection{Fermi--LAT}
\label{FermiLAT}
The Fermi Large Area Telescope (LAT) instrument is sensitive to $\gamma$ rays between energies from \R{20}\,MeV to $> 300$\,GeV \citep{atwood_large_2009,atwood_pass_2013,abdollahi_fermi_2020}. We include all events within a 10\degree \, circular region of interest \R{(ROI)} from J1405 \R{(RA, Dec) = (211\fdg2982,-61\fdg3321)} ranging from MJD 54,682 to MJD 60,492 \R{(from 2008 August 4 to 2024 July 1). We chose the energy range between 200\,MeV and 500\,GeV to avoid the large PSF associated with Fermi--LAT's lower energies and to match the original analysis in \citet{corbet_discovery_2019}. The event\RR{s} were filtered with standard data quality selections ``(DATA$\_$QUAL$>$0)$\&\&$(LAT$\_$CONFIG==1)'', and excluding zenith angles above 90\degree. We chose to include both front and back converting events using ``evtype=3'' and ``evclass=128''.} 

The Fermi--LAT events were processed using \texttt{FermiTools}, \R{ \citep[version 2.2,][]{atwood_large_2009} and the FermiPy analysis package \citep[version 1.2.0,][]{2017ICRC...35..824W} for Python}. \R{All Fermi--LAT analyses use Pass 8 data \citep{atwood_pass_2013,bruel2018fermilatimprovedpass8event} and we adopt the 4FGL-DR4 catalog} \citep{4fgldr3} source model and include all sources within \R{a larger} 15$^{\mathrm o}$ ROI to account for Fermi--LAT's PSF. \R{We derive a new orbital period of $P=13.7157\pm0.0014$\,days (see Section~\ref{sec:gamma-ray}) according to the procedure listed in \citet{corbet_discovery_2019}. This new period is in agreement with the original of $13.7135\pm0.0019$\,days, refined with more than 5\,years of additional photon events.} Close to 16 years of photon events are folded into 14 phase bins according to \RR{this} updated orbital period \R{using the same reference epoch, t$_0$, MJD 56498.7.} Each phase selection was further divided into 16 logarithmically-spaced energy bins \R{(while taking account of energy dispersion)}, \R{for spectral analysis,} using the recommended FermiTools prescription\footnote{Further Fermi--LAT data documentation and analysis details can be found at \url{https://fermi.gsfc.nasa.gov/ssc/data/analysis/software/}}.

\section{X-ray Analysis}
\label{sec:X-ray}
Due to high background, all X-ray data are \R{heavily dominated by background at soft and hard energies. At 20\,keV, NuSTAR FPMA/FPMB events are dominated by background and so are filtered \RR{to include only events} between 3--20\,keV. XMM-Newton \R{data} has evidence of \RRR{increased count rates due to }soft-\RR{proton} flaring \RRR{\citep{Weisskopf_2002, 1263836, Fioretti_2024}}, as well as heavy background at \RRR{the} soft and hard \RRR{energies} and so the events are filtered \RR{to include events} between 1--9\,keV. These energy ranges are used for all subsequent analyses.}
\subsection{Spectral Analysis}
\label{X-ray Spectra}

Spectral fitting was done using \texttt{XSPEC} (version  12.13.0c) \citep{1996ASPC..101...17A}.
We fit the observations at the Gamma-ray Maximum and the X-ray Maximum with a power-law (PL)
and an intrinsic Galactic absorption component (\texttt{tbabs}, using photoionization cross sections from \citealt{1996ApJ...465..487V} and \texttt{wilm} abundances; \citealt{wilms_absorption_2000}). The spectra from NuSTAR FPMA/FPMB and XMM-Newton MOS/PN were fitted simultaneously and tied with a constant in the model to account for the cross-normalization between the different instruments, where the cross-calibration constants were normalized to C$_\mathrm{FPMA}$. Fluxes from each instrument were calculated by including the multiplicative model component \texttt{cflux}. \RR{The b}est-fit results are shown in Table~\ref{tab:Joel Addition} along with chi-squared values \R{and degrees of freedom (DOF)}. Spectral fits for the PL model and chi-squared residuals are shown in Figure~\ref{fig:X-ray fits}. 

\begin{figure}[tbp]
    \centering
    \includegraphics[width=\columnwidth]{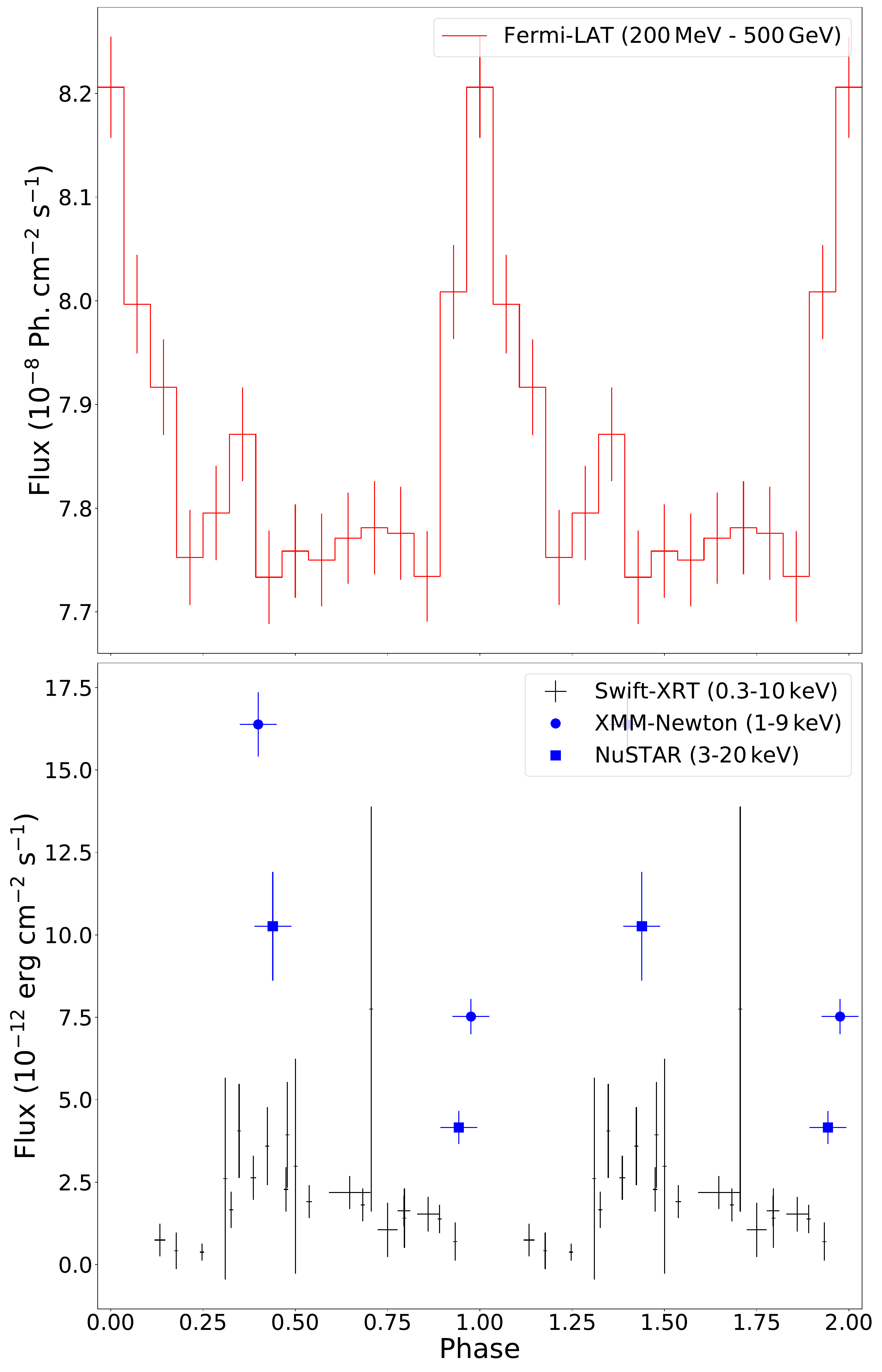}
    \caption{Gamma-ray and X-ray lightcurves folded on the orbital period from \cite{corbet_discovery_2019}. Top: Gamma-ray orbital phase folded lightcurve from Fermi--LAT. Data is taken from the probability-weighted aperture-photometry data. Bottom: X-ray orbital phase-resolved lightcurves from Swift-XRT, NuSTAR and XMM-Newton. Swift archival fluxes are shown in black. NuSTAR and XMM-Newton observations are shown in blue circles and squares respectively. A twin-peaked structure is shown at the Gamma-ray Maximum and the X-ray Maximum in the Fermi--LAT lightcurve. These phases are the focus of the 2019 NuSTAR/XMM-Newton observations.}
    \label{fig:1 XrayLC}
\end{figure}

\twocolumngrid

A comparison of joint spectral fits between the Gamma-ray Maximum and the X-ray Maximum shows a significant improvement when the $N_\mathrm{H}$ for each phased observation is fit independently ($\chi^2$/DOF = 289.79/266), as opposed to when $N_\mathrm{H}$ \RR{is} tied across phases ($\chi^2$/DOF = 297.08/267). 

\R{Due to XMM-Newton's sensitivity in the soft X-ray range, we can investigate line features.} A \R{cyclotron resonant scattering feature (CRSF)} was reported in the Gamma-ray Maximum and the X-ray Maximum XMM-Newton observations by \citet{chiu_possible_2024}, and \RR{we tested for the presence of} this feature. In addition to a PL spectral model, we added a Gaussian \RRR{absorption feature} at 2\,keV and simulated 10,000 spectra from our best-fit values using the \texttt{XSPEC} tool \texttt{simftest} to compare the $\chi^2$ fit statistics. We find that adding the Gaussian improves the fit in less than $\sim$68\% of trials. Lastly, we do find evidence of a spectral break at $\sim$3.72\R{$\pm$0.37}\,keV during the Gamma-ray Maximum ($\chi^2$/DOF = 134.14/131) but does not significantly improve upon the PL fit ($\chi^2$/DOF = 137.03/133), and is not considered in any subsequent analysis.

\begin{figure}
  \centering

  \subfigure{
    \includegraphics[width=1\columnwidth]{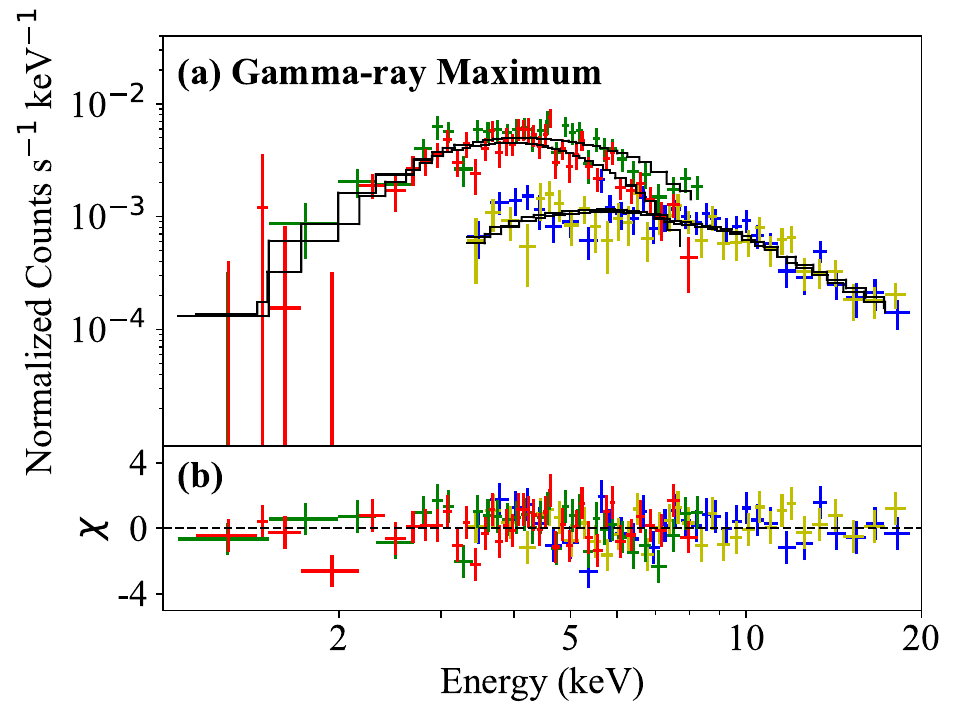}
    \label{fig:subfig1}
  }
  
  \subfigure{
    \includegraphics[width=1\columnwidth]{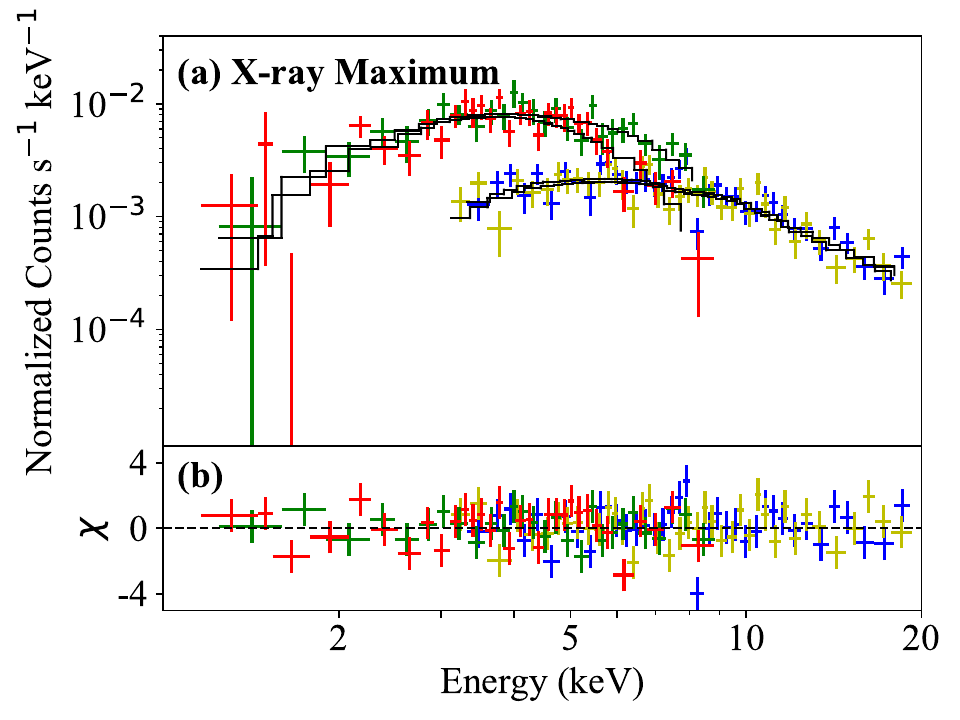}
    \label{fig:subfig2}
  }

  \caption{Joint SED fits of NuSTAR and XMM-Newton observations of J1405. The fits for each data are shown as solid lines overlaid by each instrument's data. Note that we combined MOS1 and MOS2 for \RR{better} statistics. NuSTAR FPMA/FPMB are shown in blue and gold, and XMM's combined MOS and PN are shown in green and red, respectively. The top figure shows the Gamma-ray Maximum data and its best-fit model (panel a) and its \RR{$\chi$}-squared contribution to the fit statistic (panel b). The bottom figures shows the same (best-fit model and \RR{$\chi^2$)} for the  X-ray Maximum observation.}
  \label{fig:X-ray fits}
\end{figure}

\begin{deluxetable}{cccccccc}
\tablecolumns{7}
\tablewidth{0pc}
\tablecaption{X-ray spectral parameters}
\tablehead{
\colhead{Model Parameter} & \colhead{Gamma-ray Maximum} & \colhead{X-ray Maximum}} 
\startdata
Flux$^\mathrm{a}$ &$0.57^{+0.02}_{-0.02}$ &$1.11_{-0.04}^{+0.04}$ \\
$C^b_{\rm FPMA}$ & 1 & 1\\
$C^b_{\rm FPMB}$ & 0.99\,$\pm$\,0.01 & 0.99\,$\pm$\,0.01 \\
$C^b_{\rm PN}$ & 0.90\,$\pm$\,0.07 & 0.74\,$\pm$\,0.05 \\
$C^b_{\rm MOS}$ & 1.02\,$\pm$\,0.08 & 0.80\,$\pm$\,0.06 \\
\hline
\emph{tbabs} $N_{\rm H}$  & 7.98\,$\pm$\,0.63 & 5.69\,$\pm$\,0.53 \\
$\Gamma_X$ & 1.57\,$\pm$\,0.10 & 1.42\,$\pm$\,0.08 \\
Normalization$^\mathrm{c}$ & 1.12\,$\pm$\,0.22 & 1.58\,$\pm$\,0.27 \\
\hline
$\chi^2$/DOF & 137.03/133 & 151.33/132 
\enddata
\tablecomments{\\*
$^a$ Fluxes ($\times10^{-12}$ \R{ergs cm$^{-2}$ s$^{-1}$)} integrated from 1--20\,keV. \\
$^b$ Cross Instrument Normalization. \R{$C_{\mathrm{FPMA}}$ is fixed to 1 in both the Gamma-ray Maximum and the X-ray Maximum.} \\
$^c$ $N_\mathrm{H}$ measured in $\times$10$^{22}$\,cm$^{-2}$.\\ 
$^d$ Model Normalization ($\times10^{-4}$ photons\,keV$^{-1}$ cm$^{-2}$ s$^{-1}$) at 1\,keV.}
\label{tab:Joel Addition}
\end{deluxetable}

\pagebreak
\subsection{X-ray Timing Analysis} \label{sec:X-ray timing}

We fold the observation-integrated \RR{(for easy comparison to Swift-XRT and Fermi--LAT fluxes)} XMM-Newton and NuSTAR fluxes on the binary's \RR{updated} orbital period reported and overlay the 3--10\,keV fluxes (calculated using \texttt{cflux}) onto archival Swift-XRT orbital phase-folded fluxes. We re-calculate the XRT fluxes to ensure each observation had an associated 3$\sigma$ flux error. The fluxes were extracted using the \RR{Swift--XRT data products generator webtool} \citep{2007A&A...469..379E,2009MNRAS.397.1177E} in Figure~\ref{fig:1 XrayLC}.

\R{We do not fit the X-ray lightcurves and refer to \citet{saavedra_nustar_2023}, which finds agreement with the initial results in \citet{corbet_discovery_2019} and long-term variability in the X-ray lightcurve throughout the XMM-Newton and NuSTAR observations. Instead, we focus on pulsation searches within the XMM-Newton and NuSTAR observations and modeling of the average flux state.} 

\begin{figure}[tbp]
  \centering

  \subfigure{
    \includegraphics[width=\columnwidth]{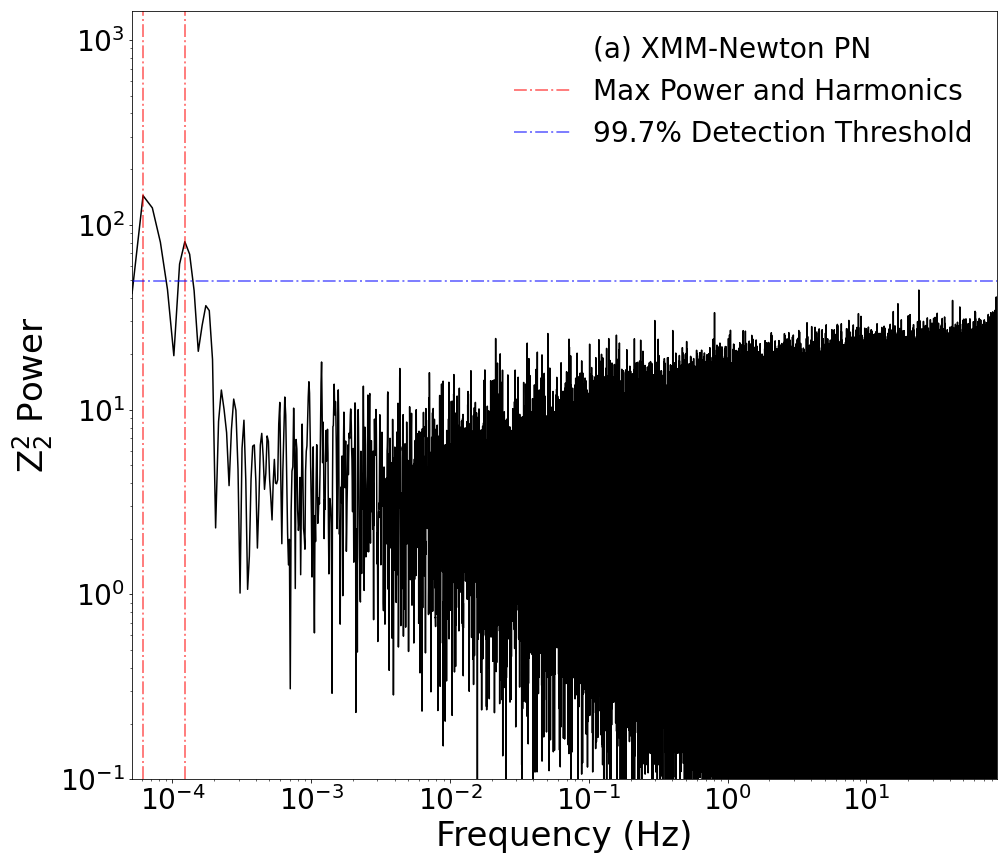}
    \label{fig:Nus_Z_2_2_0}
  }
  \hfill
  \subfigure{
    \includegraphics[width=\columnwidth]{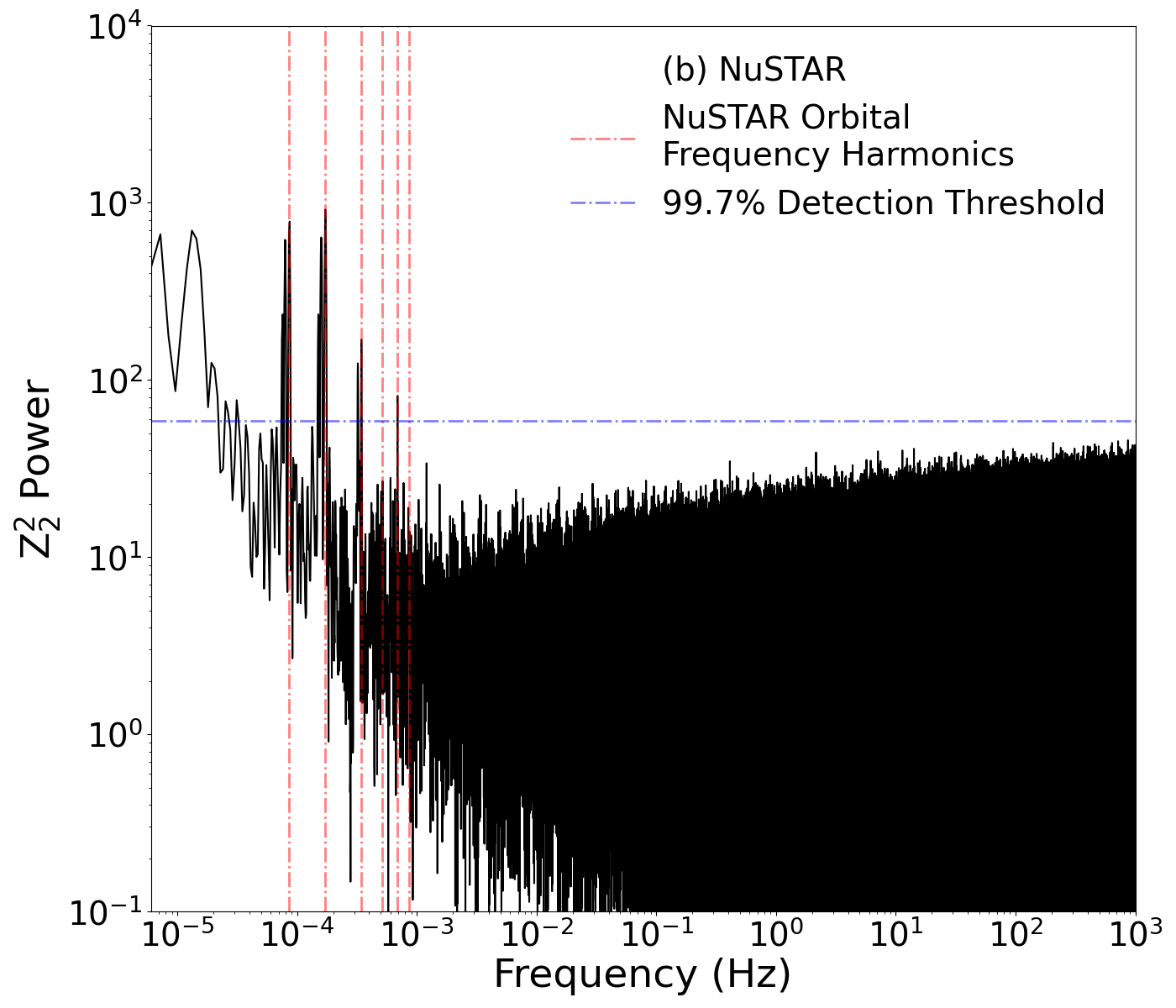}
    \label{fig:XMM_Z_2_2_5}
  }
\caption{\R{An example of two} Z$_\mathrm{2}^2$ analyses of both \R{XMM-Newton (panel (a), Gamma-ray Maximum)} and NuSTAR \R{(panel (b), X-ray Maximum)}. Only very low frequency ($1\times10^{-4}$ to $1\times10^{-5}$\,Hz) signals in panels (a) and (b) are significant besides the orbital period (and harmonics) of NuSTAR (red) in panel (b). Panel (a) shows \R{features near} 10\,Hz, however, these fail to surpass \R{a False Alarm Probability threshold (shown by the blue horizontal line at the $3\sigma$ level)}.}

  \label{fig:Z2}
\end{figure}

\RR{We searched for pulsations using the Z$_n^2$ test \citep{buccheri_search_1983} for the XMM-PN and NuSTAR FPMA/FPMB event data; we elected to use} \RRR{$n=1$ and $n=2$} provided the low signal-to-noise ratio for J1405. Barycentered NuSTAR FPMA/FPMB events were combined and analyzed separately from XMM-PN events. XMM-Newton MOS1/MOS2 were not considered due to their low time resolution. The Nyquist frequencies for XMM-Newton PN and NuSTAR are 87.72\, Hz and 7692.31\,Hz. We thus chose to search a frequency range covering  $5\times10^{-6}$\,Hz (roughly twice the observation length) to $1\times10^{3}$\,Hz and $5\times10^{-5}$\,Hz to 87\,Hz, for NuSTAR and XMM-Newton PN, respectively. \RR{We
over-sample the Z$_n^2$ tests by a factor of 5}. \R{Example
Z$_2^2$ periodicity searches} can be found in Figure~\ref{fig:Nus_Z_2_2_0} and in Figure~\ref{fig:XMM_Z_2_2_5}. We \RR{detect} no signal (ie., $>3\sigma$), however \R{features at the 1--2$\sigma$ significance level} were found throughout the Z$_{(n=1,2)}^2$ analyses. \R{Low frequency signals near $\nu=\R{1\times10^{-5}}$\,hz and below are attributed to red noise.} We do provide an upper-limit on pulse fraction\RR{s} of $\approx16\%$ and $\approx17\%$ for the Gamma-ray Maximum and the X-ray Maximum, respectively, each calculated from the XMM-Newton Z$_2^2$ analyses.
\linebreak
\linebreak

\section{Gamma-Ray Analysis}
\label{sec:gamma-ray}

\subsection{Timing Analysis}

\R{Fermi--LAT events from $200$\,MeV--$500$\,GeV \RRR{are} \RR{selected from} within a 3$\degree$ aperture \RR{of} J1405 \RRR{and} were probability-weighted using an updated source model from the 4FGL--DR4 source catalog (see Section~\ref{sec:gamma_spectral}). Weights were assigned with the fixed spectral models and the FermiTool \texttt{gtsrcprob} \citep{kerr_improving_2011}. We choose \RRR{a 3$\degree$ aperture }as it roughly matches the PSF at 200\,MeV.} Other data selections were \RR{chosen} according to the prescription detailed in \citet{corbet_discovery_2019} to provide an update with 16\,years of events (roughly 5 more years of data). \R{We conduct a power series search on the weighted aperture photometry data to refine the orbital period. We find  \RR{an orbital period of} $P=13.7157\pm0.0014$\,days, fully within the errors of the previously reported period \citep{corbet_discovery_2019}.} 

\begin{figure*}[tbp]
    \centering
    \includegraphics[width=1\textwidth]{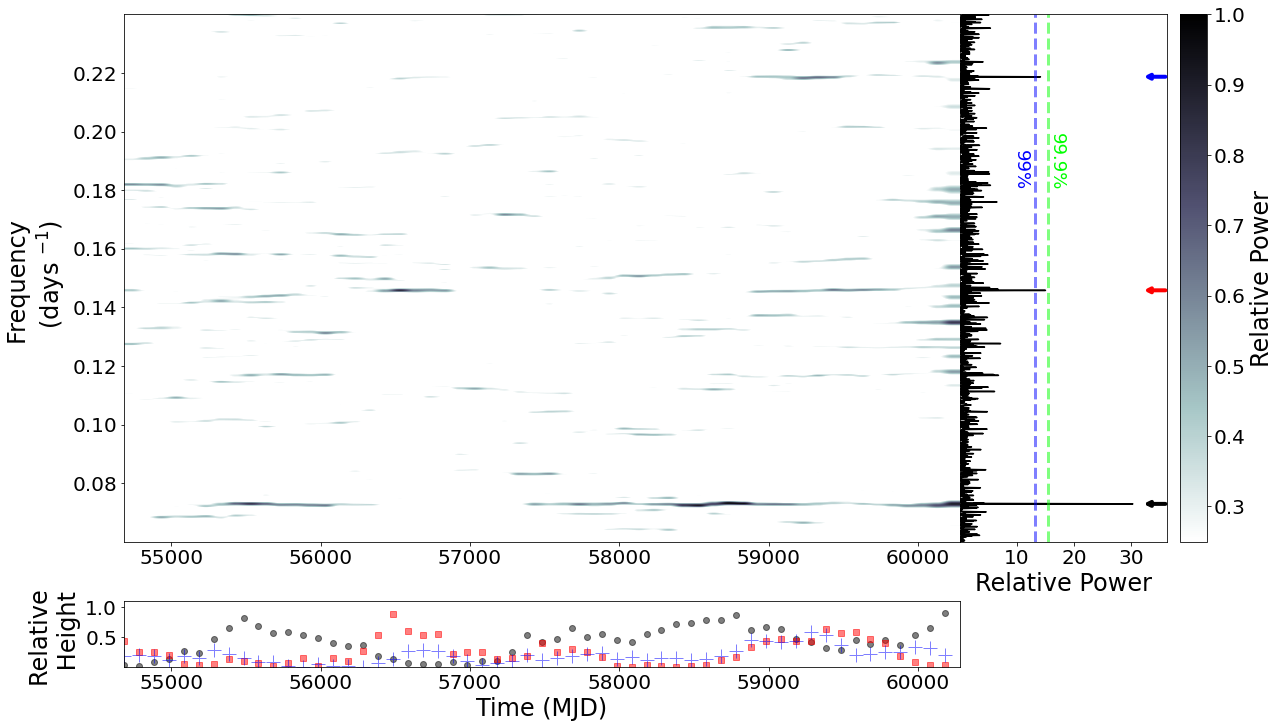}
    \caption{A dynamic Lomb-Scargle Periodogram of probability-weighted aperture-photometry Fermi--LAT data. Data were mean subtracted and weighted by the exposure. Observations are binned into 750 day bins and the \RR{relative p}ower is calculated on a 100 day width sliding window. The fundamental, 2nd and 3rd harmonic of the orbital frequency are pointed out by a \RR{black, red and blue} arrow respectively. The relative \RR{height} of the fundamental (black), first (red) and second (blue) harmonics can be seen in the bottom panel. \RR{A relative power spectrum of the full data is shown in the right panel. A dashed blue and green represent the 99\% and 99.9\% significance levels, respectively.}}
    \label{fig:DynPowSpec}
\end{figure*}

\begin{figure}
     \centering
     \includegraphics[width=\columnwidth]{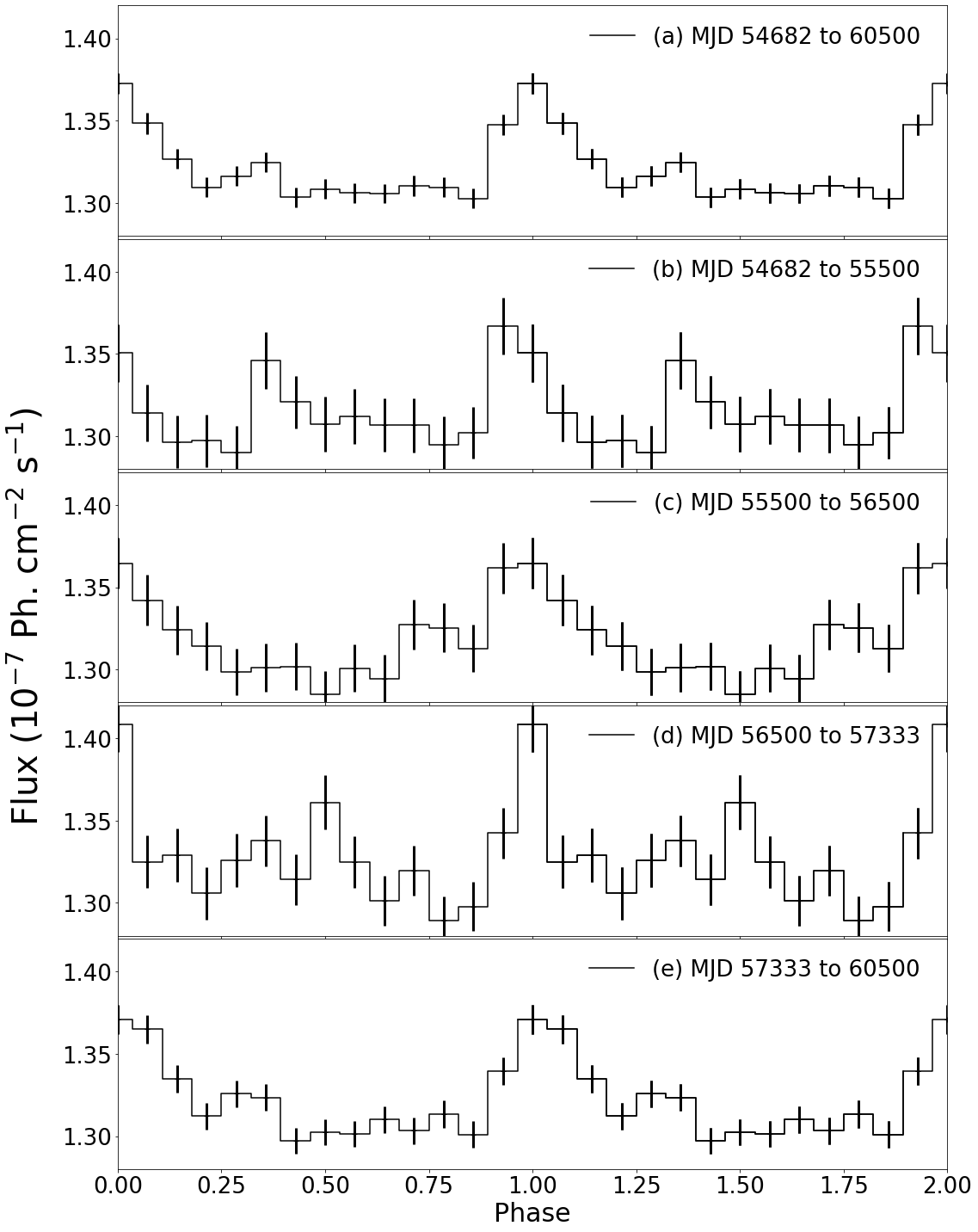}
     \caption{Probability-weighted aperture photometry from 200\,MeV--500\,GeV phase-folded lightcurves of selections between MJD 54,682--60,500. From a) to e): phase-resolved lightcurves from MJD 54,682--60,500, MJD 54,682--55,500, MJD 55,500--56,500, MJD 56,500--57,333, and MJD 57,333--60,500.}
     \label{fig:Time_res_Aperture}
 \end{figure}

\R{With an updated orbital period, we investigated the evolution of this} orbital \RR{modulation} (and its harmonics) on a dynamic timescale. We \RR{generate} 750\,day lightcurves \RR{from the Fermi--LAT events using FermiTools}. We use a dynamic Lomb-Scargle Periodogram \R{(LSP; e.g., SMC X-1; \citealt{2003MNRAS.339..447C})}, with a 100\,day sliding window function, to search for modulation between 0.06 and 0.24 days$^{-1}$. The resulting LSP, along with the relative strengths of the 2nd and 3rd orbital harmonics, are shown in Figure~\ref{fig:DynPowSpec}. \R{An inspection of the LSP shows significant strengthening and weakening of the orbital modulation, as well as a decreased significance of the 2nd harmonic. We expect these changes to account for the decreasing significance of the 2nd $\gamma-$ray peak near the X-ray Maximum between Figure~\ref{fig:Time_res_Aperture} and Figure~2 in \citet{corbet_discovery_2019}.}

We \RR{made additional time selections to search for changes in the $\gamma$-ray modulation based on the periods of relative strengths }of the orbital period and its 2nd harmonic shown in the lower panel of Figure~\ref{fig:DynPowSpec}. \R{We also fold \RR{photons on the updated binary orbital period (See Section~\ref{FermiLAT}) from} the following ranges: the full interval (panel (a)), MJD 54,682--55,500 (panel (b)), MJD 55,500--56,500 (panel (c), MJD 56,500--57,333 (panel (d)) and MJD 57,333--\R{60,500} (panel (e)) and their resulting time-resolved orbital phase-folded lightcurves are shown in Figure~\ref{fig:Time_res_Aperture}. \R{The phase-folded lightcurve for the entire time selection in panel (a), shows similar fluxes seen in Figure~2 of \citet{corbet_discovery_2019}.} The twin-peaked structure (from the orbital harmonic) of J1405 is visible in panel (b). Within Figure~\ref{fig:Time_res_Aperture}, panels (c) and (e) show that the 2nd peak structure disappears and is less pronounced when compared to panels (b) and (d).} \RR{The variable strength of the orbital modulation may indicate a super-orbital period (e.g., as seen in \citealt{2002ApJ...575..427G} for LS I +61 303); however, we do not find evidence for one in the power spectrum.}

\subsection{Spectral Analysis}
\label{sec:gamma_spectral}
\R{The original analysis of J1405 relied on aperture photometry and while a likelihood \RRR{spectral} analysis was done, the spectral parameters were frozen to the 4FGL catalog values\RRR{, allowing only the normalization to be fit} \citep{corbet_discovery_2019}. With $\sim$5 years of additional data, and an increased signal-to-noise, we \RR{conducted} a  phase-resolved likelihood analysis. }

\RR{Event files extracted using the FermiTools} were analyzed using FermiPy (version 1.2.0). Our source model consists of the \RR{G}alactic diffuse and isotropic diffuse models: ``gll\_iem\_v07.fits'' and ``iso\_P8R3\_SOURCE\_V3\_v1.txt''. A phase-integrated fit of all cataloged sources within the 15\degree \,ROI was first performed, where only each source's normalization is freed from the 4FGL--DR4's values and fit to obtain a ``baseline'' model. All sources were then frozen to this phase-integrated \RR{``baseline''} model (including the Isotropic and Galactic Diffuse model values), except J1405. J1405's spectral model, a \texttt{LogParabola} (Equation~\ref{eq:LP}),
\begin{equation}
\label{eq:LP}
    \frac{d N}{d E}=N_0\left(\frac{E}{E_b}\right)^{-\left(\alpha+\beta \ln \left(E / E_b\right)\right)}\ .
\end{equation}
is then fit, with all spectral parameters free\RR{d} \RRR{(E$_b$ is fixed to the 4FGL-DR4 value)}, using a binned likelihood analysis for each phase \RR{bin}. We do not consider other spectral models for J1405. Each fit required a minimum fit quality of 3 with the MINUIT optimizer. 

\begin{figure}[htpb]
    \centering
    \includegraphics[width=\columnwidth]{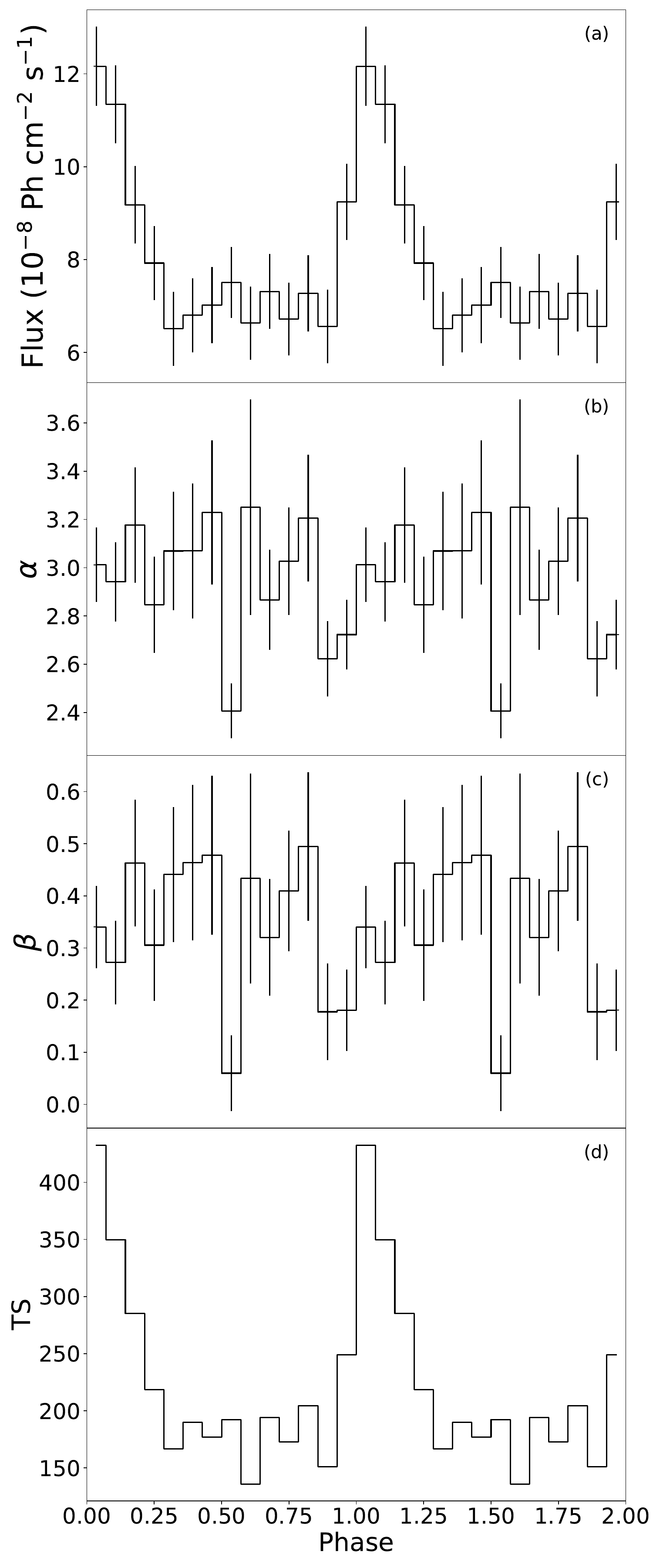}
    \caption{Phase-resolved Fermi--LAT lightcurve of J1405 using the likelihood analysis and phase evolution of spectral parameters. Each bin represents 1/14th of the binary orbit ($\sim$1\,days). Panel (a) shows the phase-resolved lightcurve showing fluxes integrated from 200\,MeV to 500\,GeV. Panels (b) and (c) show the spectral indices $\alpha$ and $\beta$. Panel (d) shows the test statistic associated with J1405.}
    \label{fig:SpectralEvo}
\end{figure}

\begin{figure}[tbp]
  \centering

  {\includegraphics[width=\columnwidth]{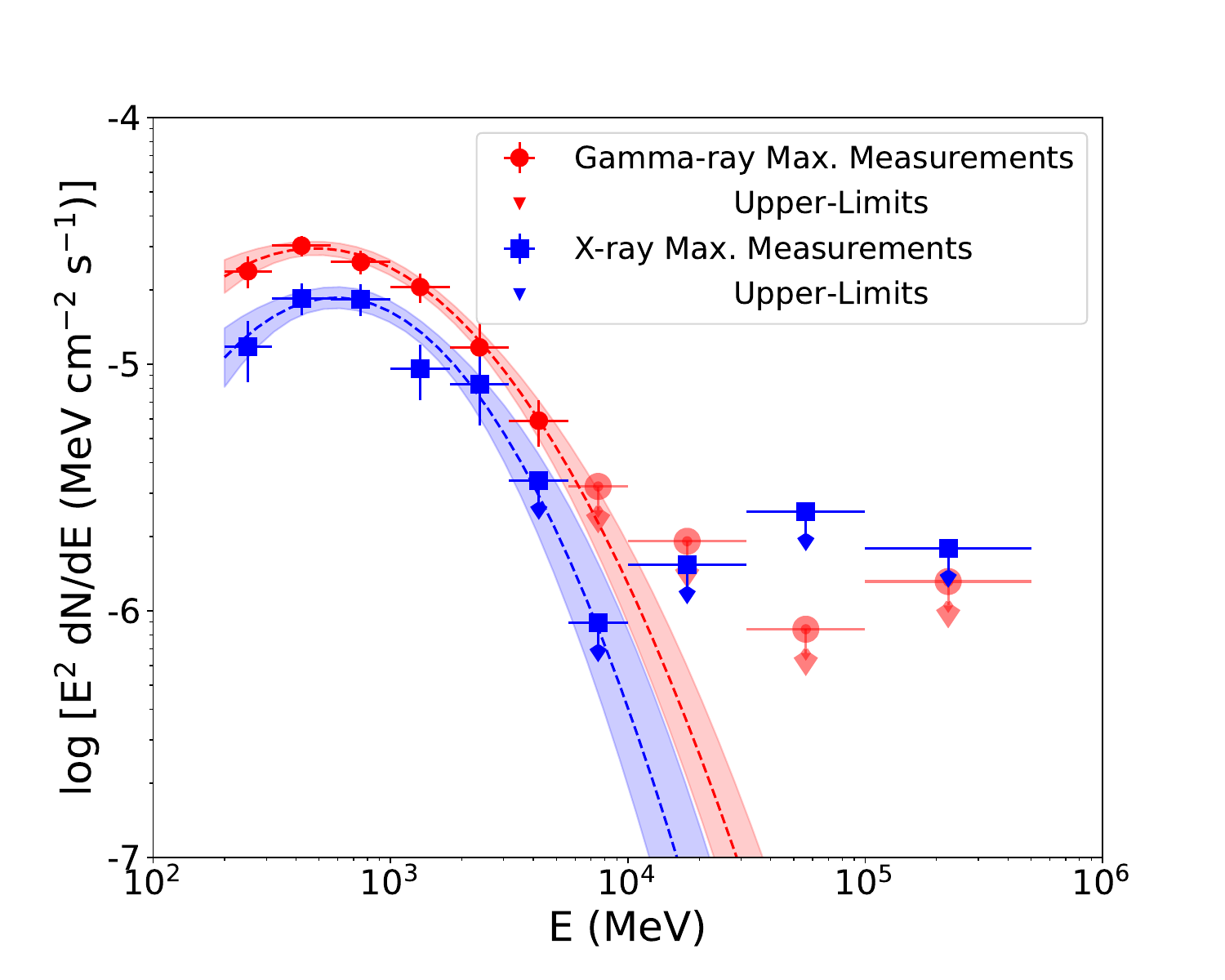}}
  \caption{Overlaid SEDs from $200\,\rm MeV-500\,\rm GeV$ are shown for both \R{the Gamma-ray Maximum (red) and the X-ray Maximum (blue)} along with their respective upper limits. The spectral models are shown as a dashed line, and 95\% confidence intervals are demarcated by the shaded regions.}
  \label{fig:FermiSEDs}
\end{figure}

The evolution of J1405's best-fit spectral curvature parameters ($\alpha$ and $\beta$) as well as the test statistic (TS; a measure of the goodness of the fit) are shown in Figure~\ref{fig:SpectralEvo} as a function of orbital phase. SEDs generated from the best-fit models \RR{were} \R{extracted from the Gamma-ray Maximum ($0.964<\phi_{\mathrm{Gamma{\text -}ray\,Max.}}<0.0357$) and the X-ray Maximum ($0.393<\phi_{\mathrm{X{\text -}ray\,Max.}}<0.464$), which overlap in phase with the XMM-Newton and NuSTAR observations,} are shown in Figure~\ref{fig:FermiSEDs}. \R{Here, we} approximate the shape of the SED at each energy bin as a power-law and allow the local spectral index to vary with \texttt{use\_local\_index=True}. \R{Additionally, spectral fluxes with a TS value $< 9$ are considered below a $3\sigma$ detection threshold, and upper limits are calculated and used instead.} A phase-resolved likelihood analysis of the two epochs between MJD 56500-57333 and MJD 57333-60500 (where there is a significant difference between the strengths of the fundamental and 2nd harmonic) found no significant changes in the spectral parameters associated with the changes in modulation.
\linebreak
\linebreak
\linebreak

\section{Broadband SED Analysis }
\label{sec:broadband}
\subsection{Theoretical Constraints}

We made the following assumptions to model the observed high-energy spectral energy distribution. The nIR spectroscopy of J1405 indicates an O6.5 III type companion star \citep{corbet_discovery_2019}, so we assume its mass, temperature and luminosity to be $\sim$35 M$_\odot$, $3.8\times10^4\,$K and $2.75\times10^5\,L_\odot$  \citep{hanson_medium_2005, mahy_spectroscopic_2015}. Using the \R{newly-derived} orbital period of \R{13.7157\,days} and this mass estimate for the O6.5 III companion, we calculated, from Kepler's 3rd law, a binary separation of $a=0.37$ AU, with the assumption that the binary orbit is circular and  the compact object is a 1.4 M$_\odot$ neutron star.

We assume the broadband SED is radiated by a population of electrons emitting synchrotron radiation and inverse Compton scattering stellar photons in a single zone close to the compact object. The stellar radiation field at the location of the compact object is then: 
\begin{equation}
\label{eq:Photon Seed Density}
    U_{\mathrm{seed}} \approx \frac{R^2}{a^2} \frac{\sigma T^4}{c} \approx 94 \, \mathrm{erg \, cm}^{-3}\ .
\end{equation}
This is much higher than the magnetic energy density $U = B^2/4\pi$ (if $B\leq 34\,\mathrm{G}$), and so, inverse Compton scattering is expected to dominate the particle energy losses. Electrons up-scatter stellar photons with an effective energy of $\epsilon \approx 2.8 kT \approx 9\,\mathrm{eV}$ \citep{moderski_2005}. The interaction occurs in the Klein-Nishina regime if the energy of the electron is such that $4\gamma \epsilon > m_e c^2$, where $\gamma$ is the Lorentz factor of the electron, and so electrons with energy $\ga 7\,\mathrm{GeV}$ ($\gamma\ga 15000$) interact in the Klein-Nishina regime. The X-ray emission is thought to be due to synchrotron emission from electrons with this typical energy since their typical synchrotron energy is $\approx 0.1\,\mathrm{keV}$ (with $B=34\,\mathrm{G}$).

The case of continuous injection of a power-law distribution of electrons cooling due to inverse Compton scattering off an external photon field and synchrotron radiation produces a specific break  associated with the transition from dominant inverse Compton cooling to dominant synchrotron cooling for the electron population \citep{moderski_2005, dubus_2006}. In the synchrotron spectrum, the \R{characteristic synchrotron energy of the electrons at E$_{\rm b}$} provides an estimate of the magnetic field that can be written as \citep{dubus_gamma-ray_2013}:
\begin{equation}
    \label{eqn:B field approx}
    B \approx \left(\frac{750 \mathrm{\,keV}}{h \nu_{\mathrm{sync}} }\right)\left(\frac{0.1 \mathrm{AU}}{a}\right)^2\,\mathrm{G}\ .
\end{equation}
Since we see no significant spectral break \RRR{with} XMM-Newton and NuSTAR, setting $E_b=h \nu_\mathrm{sync}\geq 20\,\mathrm{keV}$ implies $B\leq 2.7\,\mathrm{G}$, \RRR{and} well within the Klein-Nishina regime.

Assuming a power-law injection with $E^{-2}$ (with $E$ as the electron energy), the steady-state particle spectrum will be $E^{-3}$ for energies above the break energy, $E_{\rm b}$, and harder below $E_{\rm b}$ due to inefficient losses in the Klein-Nishina regime. The electron distribution can harden to $E^{-1.5}$ depending on how strongly inverse Compton dominates and the extent of the particle spectrum (see Figure~\ref{fig:X-ray fits} in \citealt{moderski_2005}). We chose to model this injection population with an exponential cutoff broken power law (Equation~\ref{eq:ECBPL}).

The maximum energy of the particle distribution is capped by synchrotron burnoff \citep{dejager_1996}, when the timescale to radiate the energy of the particle is equal to its gyration timescale in the magnetic field \citep{dubus_modelling_2015}.
\begin{equation}
\label{eq:energy_max}
E_{\max } \approx 60 \xi^{-1 / 2}\left(\frac{1 \mathrm{G}}{B}\right)^{1 / 2} \mathrm{TeV} \ .
\end{equation}
\R{where $\xi\geq1$ is the  acceleration efficiency, and so in the following}, we set $\xi=1$. The maximum particle energy is thus about 36 TeV (for $B=2.7\,\mathrm{G}$). 

The synchrotron spectrum is capped to a maximum around 30\,MeV with this assumption on the maximum particle energy, which is reasonable for Fermi acceleration at shocks \citep{dejager_1996}. This falls short of the GeV emission if one interprets it as synchrotron emission from the same population of electrons that radiate X-rays. Rather than assuming an arbitrary (unphysical) large maximum particle energy to enable the synchrotron emission to reach the GeV regime, we attribute the GeV component to a Maxwellian population of electrons originating from the IBS where a fraction of the pulsar wind particles are randomized instead of being accelerated to a power law  \citep{dubus_modelling_2015}.

\subsection{Broadband fitting}
\label{sec:Broadband fitting}
With these theoretical assumptions of the geometry of the binary and the contributions of synchrotron and inverse Compton emission, the broadband modeling of the X-ray and $\gamma$-ray data was conducted using the {\tt Naima} python package \citep[version 0.10][]{zabalza_naima_2015}\footnote{See technical documentation and examples at \url{https://naima.readthedocs.io/en/latest/examples.html}}, a Markov chain Monte Carlo (MCMC)  fitting package. X-ray data from XMM-Newton and NuSTAR were re-binned to contain a minimum of 100 counts per bin and \RR{we}re corrected for absorption by taking the ratio of the absorbed and unabsorbed model and applying it to each data point while taking care to include the cross-constant normalizations. The Fermi--LAT data from the Gamma-ray Maximum and the X-ray Maximum were generated from the SEDs with FermiPy, where upper-limits \RR{were} used if the TS of the energy bin \RR{was} less than 9 (corresponding to a 3$\sigma$ detection from noise). 

\begin{figure}
  \centering
  \subfigure{\includegraphics[width=0.45\textwidth]{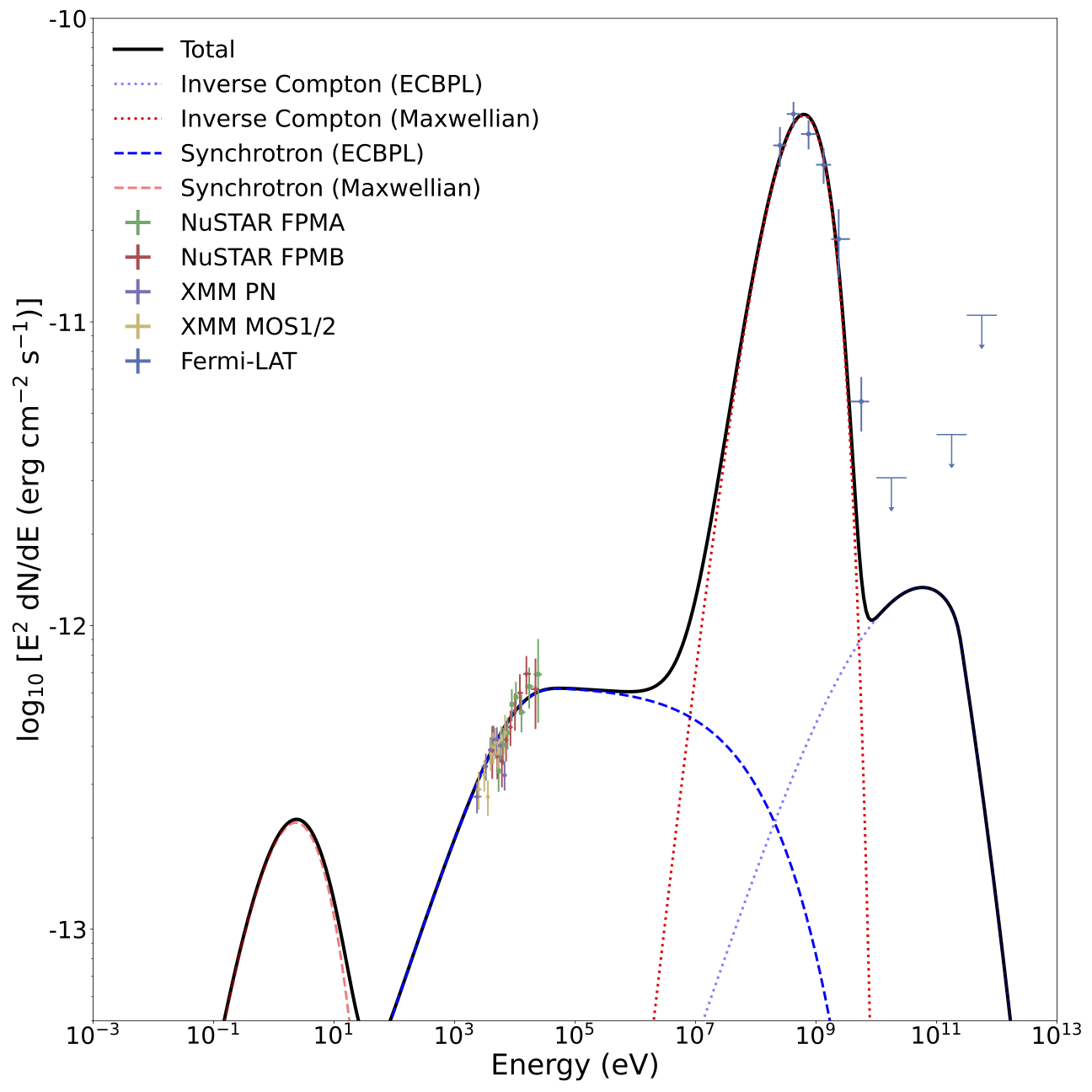}}
  \subfigure{\includegraphics[width=0.45\textwidth]{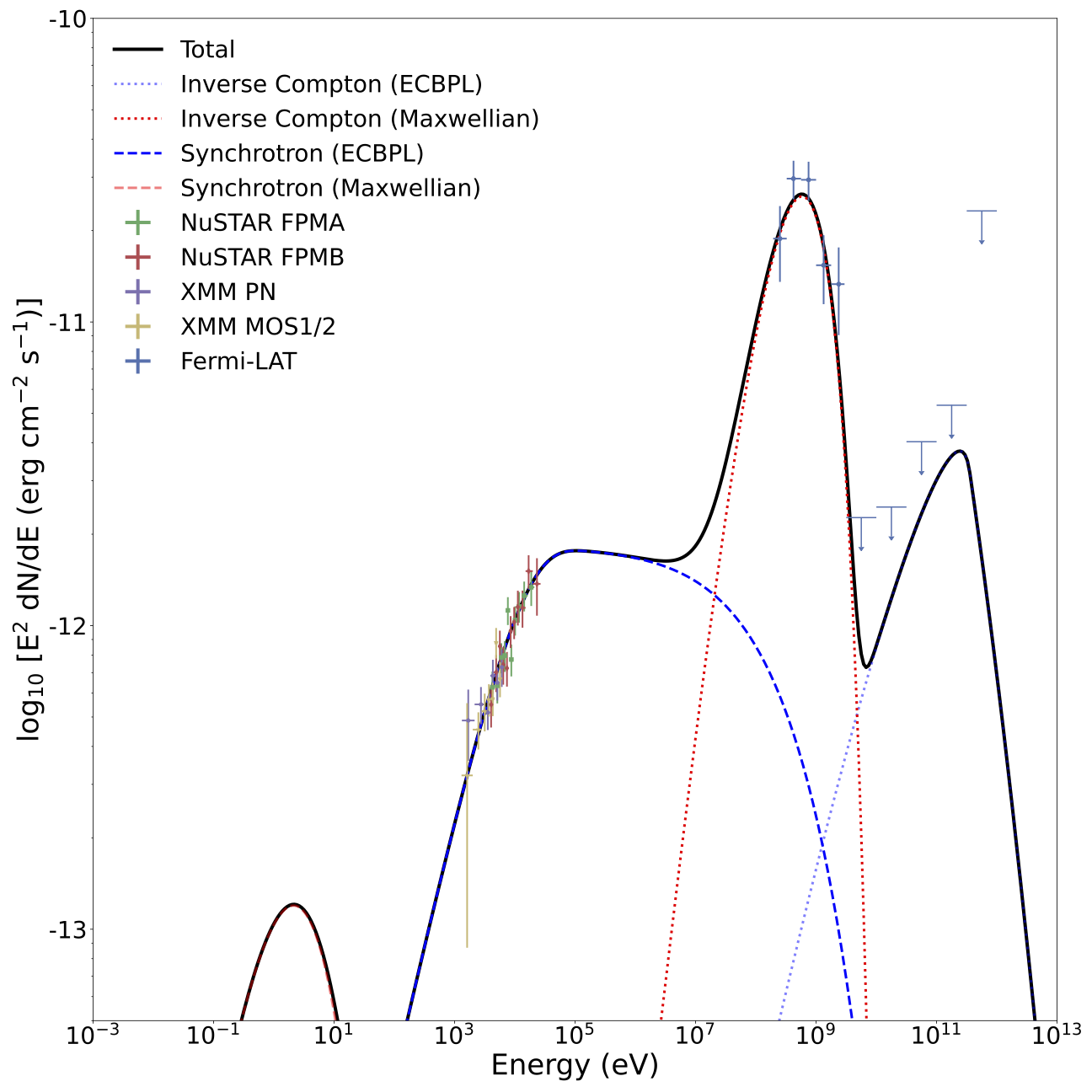}}
  
  \caption{\R{Broadband SEDs of the Gamma-ray Maximum and the X-ray Maximum. Data include XMM-Newton MOS1/2 and PN, NuSTAR FPMA/FPMB and Fermi--LAT fluxes. Fermi--LAT upper-limits represent a 95\% confidence interval and are denoted by one-sided error bars. The cumulative spectral model is shown as a black solid line. The inverse Compton components (shown as dotted lines) represent both the ECBPL (blue) and Maxwellian (red) particle distributions. Similarly, the synchrotron components are shown as dashed lines.}}
  \label{fig:BroadbandSED}
\end{figure}

Within the {\tt Naima} fitting package, as discussed above, we consider the two populations of electrons: an exponential cutoff broken power law (ECBPL, Equation \ref{eq:ECBPL})
\begin{equation}
\label{eq:ECBPL}
\begin{split}
\frac{\mathrm{d} N}{\mathrm{d} E} = \exp \left(-\left(E / E_{\text {\rm max}}
\right)\right) \times \\
\begin{cases}
A\left(E / E_0\right)^{-\alpha_1} & : E<E_{\text {b}} \\ A\left(E_{\text {b}} / E_0\right)^{\alpha_2-\alpha_1}\left(E / E_0\right)^{-\alpha_2} & : E>E_{\text {b}}
\end{cases}
\end{split}
  \end{equation}
  and a Maxwellian (Equation \ref{eq:Maxwellian}) distribution
 \begin{equation}
\label{eq:Maxwellian}
\frac{\mathrm{d} N}{\mathrm{d} E} \equiv K E^2 \exp \left(-E / E_{\mathrm{char}}\right) \ .
\end{equation}
We set $\alpha_2 = 3$, $B=2.7\rm\,G$, $E_{\rm min}=511\,\mathrm{keV}$ (the minimum energy for a $\gamma$-ray, ie.,  $\gamma=1$), $E_\mathrm{max}=36\,\mathrm{TeV}$ and $E_\mathrm{b}$ is set to vary around $1\rm\,TeV$ as per the discussion above. E$_0$ is set to 10\,TeV. We allowed $\alpha_1$, the spectral index below the break, and $E_{\rm char}$, the characteristic energy of the particles at the IBS, as well as the normalizations from both electron distributions to vary and fit using Naima. The best-fit values are given in Table \ref{tab:broadband fit results} along with their log-likelihood values (and the related $\chi^2$ values). Figure~\ref{fig:BroadbandSED} shows the broadband SEDs at the Gamma-ray Maximum and the X-ray Maximum extending from optical to TeV energies, respectively.
\RR{Using the IBS model to constrain the eccentricity, $e$, of J1405, we fitted the SED with an additional parameter,  but did not find a statistically significant estimate for 
\RRR{the eccentricity}.}
\begin{deluxetable}{ccc}
\tablecolumns{3}
\tablewidth{0pc}
\tablecaption{Broadband best-fit parameters}
\tablehead{
\colhead{Model Parameters} & \colhead{Gamma-ray Maximum} & \colhead{X-ray Maximum}} 
\startdata
ECBPL$^\mathrm{a}$ & & \\
Log$_{10}$ [A$^\mathrm{b}$ $\left(\mathrm{eV}^{-1}\right)$]& $2.98\pm0.03$ & $3.82\pm-0.04$\\
$\alpha_1$ & $1.87\pm0.02$& $1.49^{+0.02}_{-0.03}$\\
\R{E$_{\rm b} (\times 10\,\mathrm{TeV})$} & $0.26\pm0.04$ & $0.30\pm0.01$\\
\hline
Maxwellian$^\mathrm{c}$ & & \\
Log$_{10}$ [K$^\mathrm{d}$ $\left(\mathrm{eV}^{-3}\right)$]& $-5.77\pm0.04$ &$-5.59^{+0.05}_{-0.06}$\\
$E_{\rm char}$ (GeV)  & $0.77\pm0.01$& $0.60\pm0.01$ \\
\hline
LogLike & --21.95 & --16.85\\
$\chi^2 /dof$ & 43.91/35 & 33.69/34\\
\enddata
\tablecomments{\\*
$^a$ Exponential Cutoff Broken Power Law (See Equation~\ref{eq:ECBPL}) \\
$^b$ Model Normalization eV$^{-1}$ \\
$^c$ Maxwellian Distribution (See Equation~\ref{eq:Maxwellian})\\
$^d$ Model Normalization eV$^{-3}$ }
\label{tab:broadband fit results}
\end{deluxetable}

\pagebreak
\section{Discussion}
\label{sec:Disc}

\subsection{X-Ray Timing and \texorpdfstring{n$_\mathrm{H}$}{Lg}}
We find a \textgreater
$2\sigma$ detection during XMM-Newton observation 0852020101 at a frequency of $\nu\approx22$ Hz on the $Z_2^2$ search, but this detection was not consistent throughout the different instruments and observations. Therefore we report no detection of pulsations between frequencies of $5\times10^{-6}$ Hz to $1\times10^{3}$ Hz above a 1\% False Alarm Probability (as seen in Fig \ref{fig:Nus_Z_2_2_0} and \ref{fig:XMM_Z_2_2_5}) besides the orbital period of NuSTAR and its harmonics \RRR{and low frequency signals attributed to red noise}.
Since most HMGBs do not exhibit pulsations in soft \RR{($E<10\,$keV)} or hard \RR{($E>10\,$keV)} X-rays \citep{chernyakova_energy-dependent_2023}, this is not enough evidence to reject a rotation powered neutron star as the compact object. The distance ($\sim$7.7 kpc) to J1405 and strong $n_\mathrm{H}$ absorption, contribute to the low signal-to-noise and prevent pulsation searches outside the 1--20\,keV energy range in the XMM-Newton and NuSTAR observations. Although our energy range is limited, there is little precedence of finding pulsations in the soft/hard X-ray energies (e.g., PSR B1259-63 or LS I +61\degree 303; \citealt{johnston_psr_1992,weng_radio_2022}).

The simultaneous spectral fits using XMM-Newton and NuSTAR observations best-fit models at the Gamma-ray Maximum and the X-ray Maximum show two statistically different $N_{\rm H}$ column densities. At the Gamma-ray Maximum, we detected a slightly greater column density of about $(8.0 \pm 0.6)\times 10^{22}\,\mathrm{cm}^{-2}$ when compared to $(5.7 \pm 0.5) \times 10^{22}\,\mathrm{cm}^{-2}$ for the X-ray Maximum. However, a joint fit between observations does not show an improvement when allowing the Gamma-ray Maximum and the X-ray Maximum $N_\mathrm{H}$ values to fit independently over a model fixing $N_\mathrm{H}$ to the phase-averaged value. As such, we do not find a statistically significant change in the geometry of the IBS of the binary. 

We also \RR{find a} slight hardening of the X-ray photon index $\Gamma_X$ from $\sim$1.6 $\pm$ 0.1 to $\sim$1.4 $\pm$ 0.1 from the Gamma-ray Maximum to the X-ray Maximum, which is in agreement with the Swift--XRT measurements reported in \citet{corbet_discovery_2019} and the independent NuSTAR/XMM-Newton analysis in \citet{saavedra_nustar_2023}. We use the lack of a spectral break below 20\,keV as a constraint in the broadband modeling for the magnetic field (B $\approx\RR{2.7}\,$G) at the IBS and for the synchrotron radiation emission, and a more concrete limit would have implications on broadband modeling.

Previously, \RR{the presence of} a blackbody component at $\sim$1\,keV was \RR{claimed} in the X-ray spectrum observed on 2019 August 24  (Gamma-ray Maximum), making J1405 the only HMGB exhibiting a blackbody component \citep{saavedra_nustar_2023}. Moreover, \R{a CRSF} was reported in both XMM-Newton observations \citep{chiu_possible_2024} which could indicate that the compact object is a slowly rotating neutron star or a magnetar. This would classify J1405 \RR{as a source} similar to  LS 5039 \citep{yoneda_sign_2020}, although this has been refuted by \RR{\citet{volkov_nustar_2021,kargaltsev_lack_2023}}. We find that there was not a significant number of 10,000 trials ($\gtrsim 68\%$) that show an improvement to the model for either the Gamma-ray Maximum or the X-ray Maximum.  CRSF's are typically found in harder X-ray spectra \citep{staubert_cyclotron_2019}.

Furthermore, a comparison of spectral fits of the best-fit model between the full energy range of XMM-Newton and an energy range that excludes data below 2\,keV show\RR{s} no significant change in the fit quality or values of fit parameters. Due to the intrinsic high absorption, we attribute any reported feature excess below 2\,keV to be statistical noise. Both features are found below 3\,keV in the heavily absorbed X-ray spectrum where the background dominates and should be treated with \RR{caution}. Therefore, it is difficult to argue for the presence of either the cyclotron feature in the X-ray spectra. 

\subsection{Gamma-Ray Timing and Spectral Fitting}
Using \R{16} years of Fermi-LAT data, we constrain the variation of the parameters $\alpha$ and $\beta$ in the \texttt{LogParabola} spectral model. Our phase-average best-fit values of $\alpha=2.85\pm0.06$ is in good agreement with the 4FGL-DR4 catalog, however, the value for $\beta=0.38\pm0.01$, is not -- the DR4 catalog value is $0.28 \pm 0.04$. \R{This difference may be attributed to the use of likelihood weights in the DR4 catalog.}

There is a slight modulation present \RR{for the} spectral parameter $\beta$, while $\alpha$ seems to remain constant. $\beta$ appears to be anti-correlated with the $\gamma$-ray flux as shown in Figure~\ref{fig:SpectralEvo}. One bin  $\phi=0.46-0.54$ seems to have a significantly different $\alpha=2.41\pm0.11$ and $\beta=0.06\pm0.07$ and may be indicative of a physical change in the binary. However, due to the nature of the large error bars for $\alpha$ and $\beta$, it is otherwise difficult to make meaningful comments on the evolution of the SED as a function of orbital phase.

Lastly, we provide different results \RR{for the $\gamma$-ray orbital modulation compared to the} previous stud\RR{y}. In \cite{corbet_discovery_2019}, a double peaked structure was found in the $\gamma$-ray using probability-weighted photometry. However, when using a likelihood analysis, this structure at the X-ray Maximum becomes less significant during different time periods as seen in Figure~\ref{fig:Time_res_Aperture}. Upon revisiting \RR{the} probability-weighted aperture-photometry lightcurve, the structure is variable. Within the top and bottom panels of Figure~\ref{fig:SpectralEvo}, there is no significant change in the observed flux from J1405 from \citet{corbet_discovery_2019}'s discovery analysis besides the decreased feature at the X-ray Maximum ($\phi \approx 0.4$). Additionally, the dynamic power spectrum of the aperture photometry in Figure~\ref{fig:Time_res_Aperture} shows that the significance of the orbital period harmonics (the 2nd peak) \RR{varies}. This suggests periods of $\gamma$-ray activity that may be caused by changes \RR{such as} the structure of the IBS or clumps in the stellar wind \citep{kefala_modeling_2023,paredes-fortuny_simulations_2015,johnston_psr_1992}. Additionally, some supergiant high-mass X-ray binaries (sgHMXBs) have been observed to display similar variation in strength of the super-orbital period harmonics (e.g., \citealt{islam_investigating_2023}). Many of these sgHMXBs are thought to have a changing wind geometry such as a co-rotating interaction regions structures, possibly linking the two classes of binaries.

\subsection{Comparison Between IBS and microquasar Broadband Models}
By setting the spectral break in the X-ray spectrum above 20\,keV, we approximate the magnetic field to be $\sim $2.7\,G, and fit a two component IBS model similar to LS 5039 \citep{dubus_modelling_2015}. Within the ECBPL distribution, upon fixing $\alpha_2 = 3$, we find that the electron distribution indeed hardens to nearly 1.5 for the X-ray Maximum but remains softer for the Gamma-ray Maximum. Additionally, the contribution of the ECBPL distribution accounts for the increase in X-ray flux. In both cases, synchrotron emission from the ECBPL sufficiently models the X-ray spectra. As \RR{shown} in Section~\ref{sec:gamma-ray}, there is little modulation in the $\gamma$-ray SEDs and so it is sensible that the Maxwellian distribution shows little change in K and E$_{\mathrm{char}}$. The synchrotron emission from the Maxwellian distribution is in the optical band, of which, the companion's emission \RR{dominates} the spectrum. The IC emission accounts for the MeV--\,GeV emission well. However, the upper-limits of the GeV emission are best fit with the inclusion of the IC emission from the ECBPL for both Phases. Additionally, preliminary observations from the High Energy Stereoscopic System \citep[H.E.S.S.,][]{noauthor_iaat_nodate}\footnote{These H.E.S.S. observations are from a Master's thesis that 
can be found at: \\\url{http://astro.uni-tuebingen.de/publications/diplom.shtml}.} indicate VHE fluxes consistent with Fermi--LAT's upper-limits and our modeling of the GeV/TeV emission.

The microquasar and IBS models offer different mechanisms to explain the high-energy emission in HMGBs. In the microquasar model, the $\gamma$-ray emission arises from relativistic jets from the compact object as it accretes matter from its massive companion. Conversely, the IBS model focuses on the collision between the stellar winds of the binary and the PWN of the neutron star, leading to the acceleration of particles and the generation of $\gamma$ rays.

\RR{The IBS model has an advantage over the microquasar jet model in that it requires significantly fewer fitting parameters. Given the quality of the X-ray and $\gamma$-ray data, fitting a model with many free parameters would likely lead to degenerate solutions and large uncertainties, especially without direct observational evidence for components like disk or coronal emissions.}

With a few simplifications, we are able to decrease the number of free parameters to five, fitting just the normalizations of particle distributions, ECBPL slope $\alpha$, the break energy of the IBS E$_{\mathrm{b}}$, and the cutoff energy for the GeV emission $E_{\mathrm{char}}$. Additionally, because we used \texttt{Naima} and an MCMC methodology in our fitting, we find a low correlation and variance between parameters (\R{see Appendix~\ref{app:corner-phase0} and Appendix~\ref{app:corner-phase6})}. We find statistically satisfying fits of $\chi_\nu^2=1.25$ and $\chi_\nu^2=0.99$ (for the Gamma-ray Maximum and X-ray Maximum, respectively) for the broadband data with physical parameters used in previous IBS modeling of HMGBs \citep{dubus_modelling_2015}. In this interpretation, orbit-to-orbit variability could be due to fluctuations in the IBS position or properties due to stellar wind variability or clumping.

\section{Conclusion}
\label{sec:Conc}
We present detailed orbital phase-resolved multi-wavelength observations of the High-Mass Gamma-Ray Binary 4FGL J1405.1-6119, utilizing XMM-Newton and NuSTAR observations at the Gamma-ray Maximum and the X-ray Maximum (at the $\gamma$-ray maximum and minimum, respectively) supplementing nearly 16 years of Fermi--LAT data. We found no significant pulsations in the X-ray observations after searching a frequency range of roughly $1\times10^{-5}$ Hz to $1\times10^{3}$ Hz. 

X-ray spectral analysis between 1.0 and 20\,keV reveals \RR{no evidence} of \RR{either a} thermal component below 2\,keV, \RR{nor} a CRSF component at $\sim 2$\,keV\RR{. There is a} slight hardening \RR{of the} X-ray spectrum from the Gamma-ray Maximum to the X-ray Maximum, without \RR{evidence} a spectral break \RR{during} either observation. 

Remarkably, the X-ray spectrum does not exhibit softening \RR{up} to at least 20\,keV. The lack of a significant detection of a spectral break below 20\,keV allows for a rough estimate of the magnetic field at the IBS to be at most 2.7\,G\RR{, which} is used for broadband analysis.

Orbital phase-resolved analysis of Fermi--LAT data (200\,MeV -- 500\,GeV) provides the evolution of the spectral shape as a function of orbital phase. We find slight evolution of spectral parameter $\beta$, however the SED experiences very little modulation in other parameters and errors are large. The previously reported 2nd peak at $\phi\approx 0.4$ seems to fade in significance for our binned likelihood analysis and is variable during our probability-weighted aperture photometry analysis for certain time ranges.

We lastly provide an alternative broadband modeling of the X-ray and $\gamma$-ray data to the microquasar model with an intrabinary shock model to explain the high-energy emission. Our IBS model provides a good fit with fewer free parameters and results in $\chi^2_\mathrm{red}$ values of $\sim$ 1.25 and $\sim$ 0.99 for both the Gamma-ray Maximum and the X-ray Maximum. 

Due to the high absorption of the soft X-rays, and discovery of the weakening $\gamma$-ray modulation, we recommend further (and more sensitive) X-ray and GeV and TeV $\gamma$-ray observations to better understand the nature and high-energy emission of 4FGL J1405.1-6119.

\medskip\noindent{\bf Acknowledgments:}
\RR{The authors thank the referee for useful comments and feedback. }The Fermi--LAT Collaboration acknowledges generous ongoing support
from a number of agencies and institutes that have supported both the
development and the operation of the LAT as well as scientific data analysis.
These include the National Aeronautics and Space Administration and the
Department of Energy in the United States, the Commissariat \`a l'Energie Atomique
and the Centre National de la Recherche Scientifique / Institut National de Physique
Nucl\'eaire et de Physique des Particules in France, the Agenzia Spaziale Italiana
and the Istituto Nazionale di Fisica Nucleare in Italy, the Ministry of Education,
Culture, Sports, Science and Technology (MEXT), High Energy Accelerator Research
Organization (KEK) and Japan Aerospace Exploration Agency (JAXA) in Japan, and
the K.~A.~Wallenberg Foundation, the Swedish Research Council and the
Swedish National Space Board in Sweden.

This work was also partially supported by NASA
grant 80NSSC20K1304 and also under NASA award number
80GSFC21M0006.
Based on observations obtained with XMM-Newton, an ESA science mission with instruments and contributions directly funded by ESA Member States and NASA. This research has made use of data from the NuSTAR mission, a project led by the California Institute of Technology, managed by the Jet Propulsion Laboratory, and funded by the National Aeronautics and Space Administration. Data analysis was performed using the NuSTAR Data Analysis Software (NuSTARDAS), jointly developed by the ASI Science Data Center (SSDC, Italy) and the California Institute of Technology (USA). This work made use of data supplied by the UK Swift Science Data Centre at the University of Leicester.
\pagebreak
\appendix
\restartappendixnumbering
\renewcommand\theequation{A.\arabic{equation}}
\renewcommand\thefigure{A.\arabic{figure}}    

\section{Broadband Modeling: MCMC Corner Plots}

From the broadband modeling with the \texttt{Naima} python module, we generate corner plots to show the variance and correlations between the fit parameters from the ECBPL and Maxwellian distributions listed in Table~\ref{tab:broadband fit results}. These corner plots show the 2-dimensional distributions of posteriors and the maximum likelihoods for each set of parameters. On the diagonal, 1-dimensional distributions are also shown for each parameter. For both the Gamma-ray Maximum and the X-ray Maximum, we find no strong correlations and no bi-modal distributions of parameters. 

\begin{figure}[ht!]
    \centering
    \includegraphics[width=0.75\linewidth]{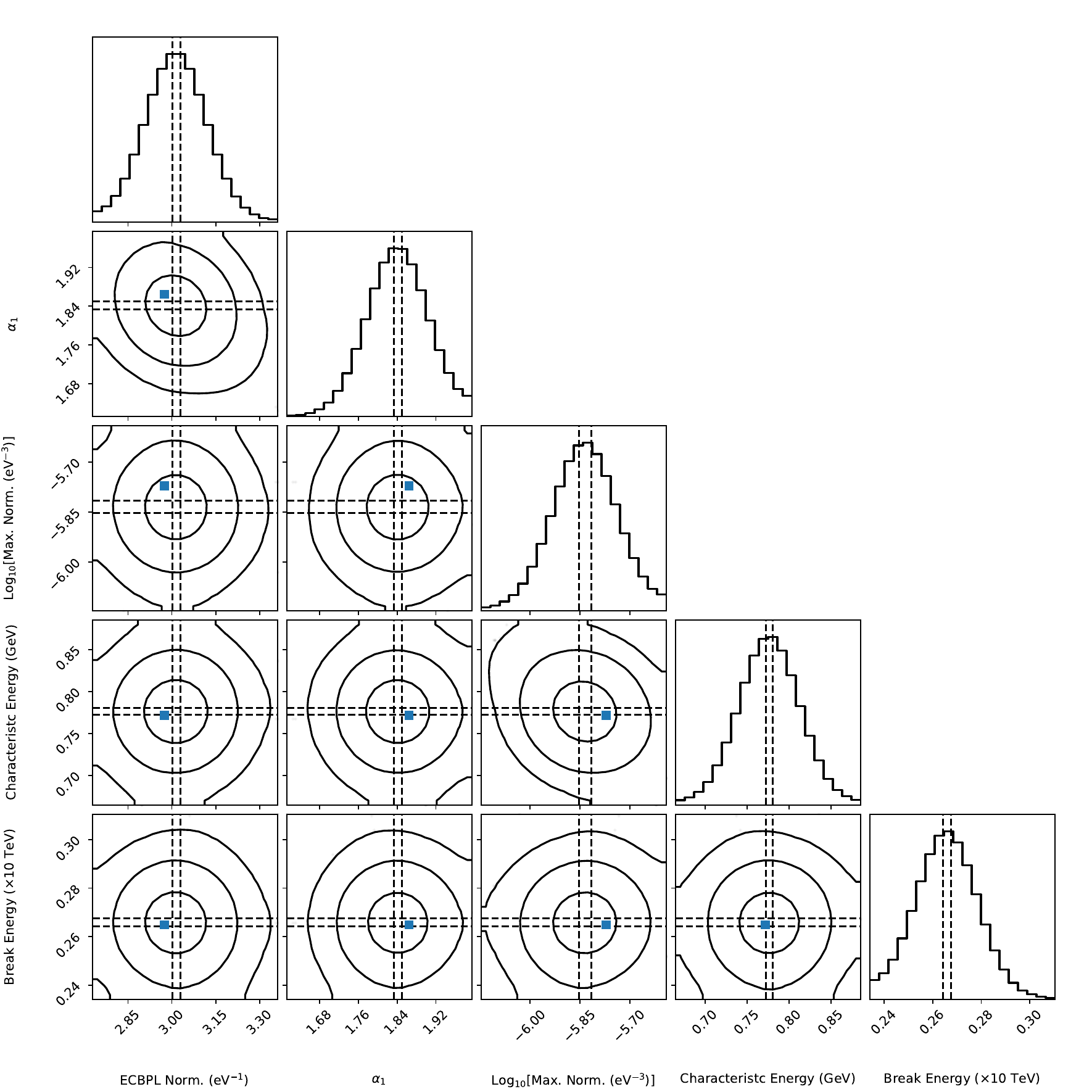}
    \caption{A corner plot of the sample density in the two dimensional parameter space of all parameter pairs of the broadband fit of the Gamma-ray Maximum. Confidence intervals of 1, 2 and 3$\sigma$ for a 2-dimensional Gaussian distribution are overlaid for each panel, as well as the 1-dimensional 68\% confidence interval along the diagonal. Dashed horizontal and vertical black lines indicate the 1-dimensional 68\% confidence interval for each parameter. The maximum likelihood value for each parameter is shown in blue.}
    \label{app:corner-phase0}
\end{figure}
\begin{figure}[ht!]
    \centering
    \includegraphics[width=0.75\linewidth]{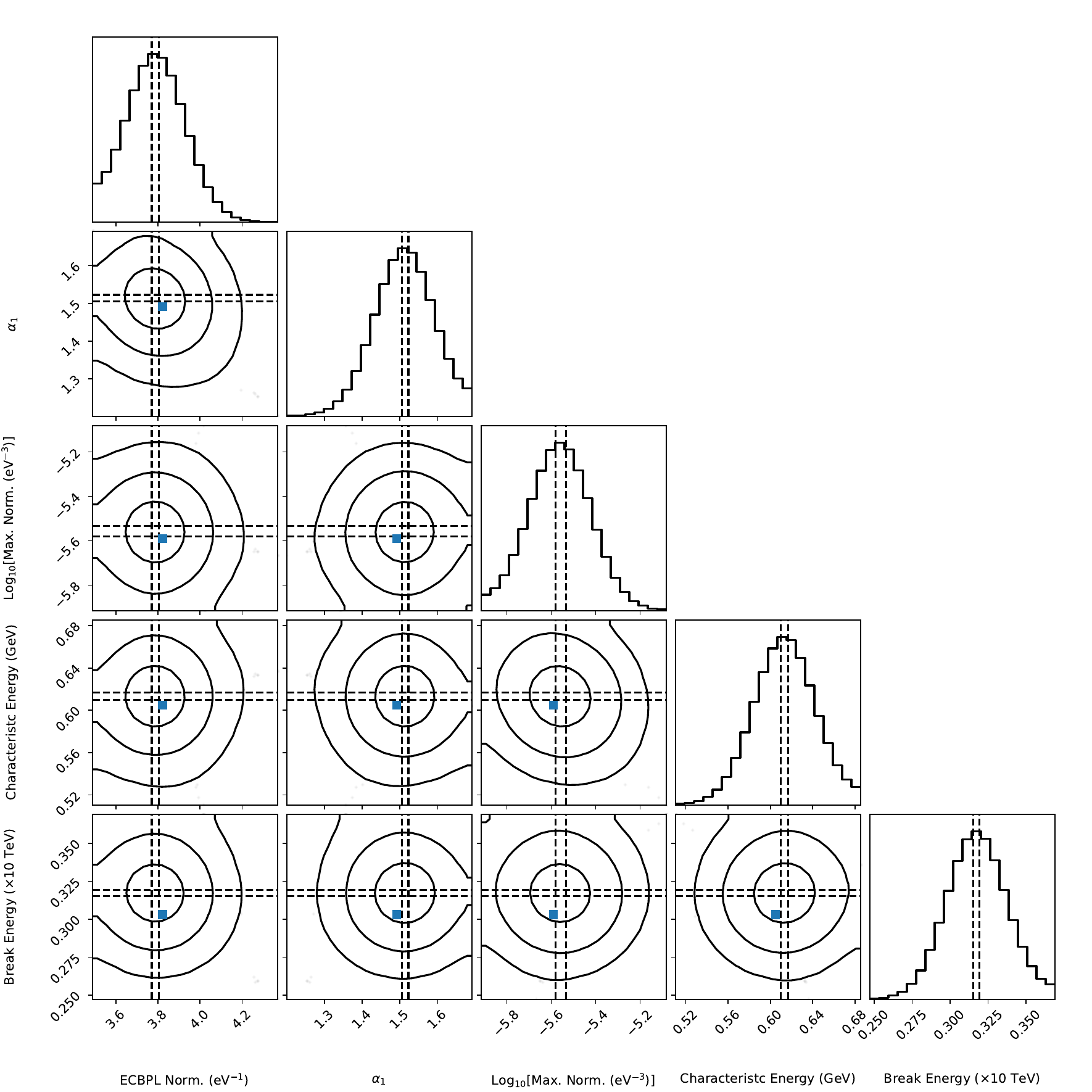}
    \caption{As in Figure~\ref{app:corner-phase0}, but for the X-ray Maximum fit.}
    \label{app:corner-phase6}
\end{figure}

\clearpage
\vspace{5mm}

\pagebreak
\pagebreak
\bibliography{bibtex}{}
\bibliographystyle{aasjournal}
\end{document}